\pdfoutput=1

\documentclass[a4paper, 10pt, reqno]{amsart}
\usepackage{packages}

\setcitestyle{semicolon}

\newcommand{\orcid}[1]{\hspace{2pt}\orcidlink{#1}\hspace{2pt}}

\newcommand{\citeprefix}[2]{\cite[#1][]{#2}}

\newtheorem{theorem}{Theorem}[section]

\newtheorem{proposition}[theorem]{Proposition}

\title{%
        The Morse Transform for Discrete Shape Analysis
        }

\author{Alexander M.~Tanaka\orcid{0000-0001-5769-5497}\textsuperscript{\faCoffee}
}
\thanks{%
\textsuperscript{\faCoffee}
Mathematical Institute, University of Oxford, Woodstock Road, OX2 6GG, UK}

\author{Aras T.~Asaad\orcid{0000-0001-9578-8838}\textsuperscript{\faLaptopMedical}
}
\thanks{%
\textsuperscript{\faLaptopMedical}
Oxford Drug Design Ltd, Oxford Centre for Innovation, 9 Alfred St, Oxford, OX1 4EH, UK}

\author{Richard I.~Cooper\orcid{0000-0001-9651-6308}\textsuperscript{\faLaptopMedical, \faFlask}
}
\thanks{%
\textsuperscript{\faFlask}
Inorganic Chemistry Laboratory, University of Oxford, South Parks Road, Oxford, OX1 3QR}

\author{Vidit Nanda\orcid{0000-0001-9243-6749}\textsuperscript{\faCoffee, \faEnvelope[regular]}
}
\thanks{%
\textsuperscript{\faEnvelope[regular]}
Corresponding author. E-mail: \href{mailto:nanda@maths.ox.ac.uk}{nanda@maths.ox.ac.uk}}

\begin{document}

\begin{abstract}
The geometry of an object plays a vital role in modulating its interactions with the physical world. 
It nevertheless remains difficult to describe geometric information numerically for the purposes of statistical inference or classification tasks. 
Here, we introduce a new topological transform which leverages directional piecewise-linear Morse theory to quantify the geometry of an embedded object by cataloguing critical points across multiple height-functions. 
The output of this Morse transform records both the heights and the local topological type (peak, trough or saddle) of the critical points that characterise the underlying shape, retaining finer information than the Euler characteristic transform whilst naturally prioritising a shape's outermost regions. 
Crucially, this output can be further compressed into a rich but compact feature vector. 
We benchmark the Morse feature vector as a descriptor for ligand-based virtual screening (LBVS), which intrinsically depends on the shape of molecules. 
Under a common gradient-boosted tree classification pipeline, Morse descriptors achieve the highest mean AUROC when compared to other topological transform descriptors and to standard shape-based LBVS descriptors. 

\end{abstract}

\maketitle

\markleft{\MakeUppercase{Alexander M. Tanaka, Aras T. Asaad, Richard I. Cooper, and Vidit Nanda}}

\section{Introduction}
\lettrine[lines=2, loversize=0.01, findent=0pt, nindent=0pt]{T}{opological} data analysis \citeprefix{TDA;}{roadmap} analyses the shape of high-dimensional data by examining it at multiple scales and extracting computable topological invariants. By far the best-known of these invariants is persistent homology \citep{persistence}, which assigns lifespans to topological features (such as connected components, tunnels and cavities) across nested sequences of spaces. Applications abound, ranging from the study of
phase transitions \citep{tda_analysis_of_phase_transitions_2016} to
rock pore geometry characterisation \citep{tda_rock_pore_geometry_characterisation_2018}, 
traffic pace patterns \citep{tda_traffic_analysis_2021}, 
neuronal tauopathy \citep{tda_neuronal_tauopathy_2023},
audio identification \citep{tda_audio_identification_2024} and beyond. 

Directional topological transforms, such as the persistent homology transform and Euler characteristic transform \citep[PHT and ECT;][]{pht}, extend this paradigm further by examining the Euler characteristic and persistent homology of subspace sequences constructed by changing height thresholds across various directions. 
The payoff is an injectivity result for generic shapes -- if two embedded simplicial complexes $X$ and $Y$ have $\text{ECT}(X) = \text{ECT}(Y)$, then $X$ and $Y$ coincide \citep{pht_how_many_directions, pht_injectivity_ghrist}. 
Variants of the ECT have been applied in predicting clinical outcomes in glioblastoma \citep{sect}, morphology of barley seeds \citep{ect_barley_seeds} and time-varying shape data \citep{detect_ect}. 

Ordinary persistent homology and its directional avatar are underpinned by Morse theory \citep{morse_milnor}. 
The core mantra of Morse theory is that one may reconstruct (up to homotopy type) a smooth manifold $M$ from a generic smooth map $f:M \to \mathbb{R}$ by constructing a cell complex with one cell per critical point of $f$. 
The attaching maps between these cells can be
rather intricate, so one often seeks solace in recovering weaker topological invariants of $M$, such as homology or Euler characteristic, from the \textit{local Morse data} of a given $f$. 
Likewise, discrete variants of Morse theory \citep{kosinski1962, brehm_kuhnel_1987, pl_bestvina_brady, forman_1998, pl_morse_low_dim} provide an efficient mechanism for simplifying  (persistent) homology computation of (filtered) simplicial complexes
\citep{mn, hmmn}.

\subsection*{Contributions}

In this work, we define a new topological transform, the \textit{Morse transform}, based on piecewise-linear Morse theory to efficiently summarise the shape of a simplicial complex $K$ embedded in Euclidean space $\mathbb{R}^n$. 
This transform is indexed by an integer depth parameter $d \geq 0$; at depth $d$, the Morse transform assigns to each unit vector $\xi \in \mathbb{R}^n$ a matrix whose rows correspond to the top $d$ critical points of the height function prescribed by $\xi$ and whose $(n+1)$ columns tabulate the relevant local Morse data at that critical point. 

The Morse transform distinguishes itself from other directional topological \linebreak transforms by: (i) providing more explicit local information than the ECT or the PHT; (ii) employing a finer topological invariant than the ECT; and (iii) prioritising the outermost regions of discrete shapes, which are the most relevant for analysing surface geometry. 
At maximal depth, the Morse transform can recover the ECT;
thus, the maximal depth Morse transform inherits its injectivity from the $\text{ECT}$.

In order to make the Morse transform amenable to downstream machine learning tasks, we describe a vectorisation scheme to compress its output to a single vector whilst retaining much of its rich structure. The length of this compressed vector depends on the ambient dimension $n$, but remains agnostic to 
the depth $d$,
the number of directions along which the Morse transform has been evaluated 
and
how these directions are labelled. 
We conduct a comprehensive empirical sensitivity study of the Morse shape descriptor, demonstrating that it is well-behaved with regard to fluctuations in the number of directions and depth, as well as being robust to perturbations of directions and positional noise.

\subsection*{Benchmarking}

We benchmark the Morse transform on \textit{virtual screening} datasets \citep{brogi2019computational} for three reasons: first, it is an exemplar problem in which geometry is known to drive a vital physical property: protein–ligand binding is governed largely by molecular shape complementarity \citep{shape_and_VS}. 
Second, the Euler characteristic transform and its relatives have already been validated as effective shape classifiers across a range of datasets \citep{sect, ect_barley_seeds}; what remains open is whether or not the finer information retained by the Morse transform translates into an advantage. 
Third, we aim to establish not only that the Morse transform can classify geometry, but also that its compressed vectorisation can combine successfully with non-shape features in a machine-learning pipeline.

We show here that the Morse feature vector achieves the highest AUROC in comparison to
ECT-based descriptors and specialised shape-based LBVS descriptors when classified with the same gradient boosted tree model on
a series of virtual screening tasks. 
The use of TDA features has recently become quite popular in biomolecular property prediction \citep[see for instance][]{topologynet_gww_2017, ph_cang_gww_2018, agl_score_gww_2019, psh_ml_2021, todd_2022, tda_llm}. 
To facilitate the use of directional topological transforms in such pipelines, we provide a codebase to implement the Morse transform for 3D molecular simplicial complexes. 
We further implement our Morse transform-based LBVS method; the core steps of our pipeline form an adaptable blueprint for using the Morse transform for general discrete shape analysis.

\subsection*{Outline} 

In Section~\ref{sec:prelim} we establish the mathematical preliminaries and notation that form the foundation of the rest of the paper; 
in Section~\ref{sec:morse_transform} we unveil the Morse transform for analysing simplicial complexes; 
in Section~\ref{sec:methodology} we describe how to generate molecular simplicial complexes and their Morse transform-based features; 
in Section~\ref{sec:experiments} we present the performance of our Morse transform-based LBVS method against other topological transforms and LBVS methods on virtual screening tasks; 
and finally we conclude in Section~\ref{sec:conclusion}.

\section{Preliminaries}
\label{sec:prelim}

\subsection{Simplicial homology}
\label{prelim:homology}

To each finite simplicial complex $K$ of dimension $m$, one can associate a sequence of vector spaces called \textit{(reduced) homology} groups \citeprefix{Ch.~2 of}{hatcher} as described below. We impose an arbitrary ordering on the vertices, so that each $k$-simplex $\sigma$ is uniquely expressible as a tuple of vertices $(v_0,v_1,\ldots,v_k)$ written in ascending order. The faces of $\sigma$ inherit this ordering: for each $i$ in $\{0,\ldots,\dim \sigma\}$, we let $\sigma_{-i}$ denote the face of $\sigma$ obtained by removing $v_i$. 

We write $C_k$ for the real vector space obtained by treating the $k$-simplices of $K$ as a basis. The $k$-th \textit{boundary operator} is the linear map $\partial_k:C_k \to C_{k-1}$ whose action on a basis $k$-simplex $\sigma \in K$ is given by $\partial_k(\sigma) = \sum_{i=0}^k (-1)^i\sigma_{-i}$. Let us also define $\partial_{0}:C_0 \to \mathbb{R}$ as the map which sends each basis vertex to $1$. Thus, we have a descending sequence of vector spaces and linear maps:
\[
\cdots 
\stackrel{\partial_{k+1}}{\longrightarrow} C_k 
\stackrel{\partial_k}{\longrightarrow} C_{k-1} 
\stackrel{\partial_{k-1}}{\longrightarrow} 
\cdots 
\stackrel{\partial_1}{\longrightarrow} C_0 
\stackrel{\partial_0}{\longrightarrow} \mathbb{R} 
\stackrel{\partial_{-1}}{\longrightarrow}0
\:.\]
A routine calculation confirms that the composite $\partial_k \circ \partial_{k+1}$ is the zero map for all $k \geq -1$; or equivalently, the kernel of $\partial_k$ contains the image of $\partial_{k+1}$. The quotient vector space $\widetilde{H}_k(K) \coloneq \ker \partial_k/\text{img }\partial_{k+1}$ is called the $k$-th \textit{reduced homology group} of $K$; and its dimension, denoted $\widetilde{\beta}_k(K)$, is called the $k$-th \textit{reduced Betti number} of $K$. Homology groups satisfy several remarkable properties, including homotopy-invariance, functoriality and efficient computability. For our purposes here, it suffices to note that the reduced Betti numbers of the one-point space are all zero and that $\widetilde{\beta}_{-1}(X)$ vanishes if and only if $X$ is nonempty.
We say that $K$ is \textit{acyclic} whenever $\widetilde{\beta}_k(K) = 0$ holds for all $k$.

\subsection{Piecewise-linear Morse theory}
\label{prelim:plmorse}

$K \subset \mathbb{R}^n$ be a finite simplicial $n$-complex and denote
its $n$-skeleton by $K_n = \{\sigma \in K \mid \dim \sigma \leq n \}$.
An assignment $f:K_0 \to \mathbb{R}$ is called a \textit{piecewise-linear} (PL) \textit{Morse function} on $K$ if it is injective on every simplex \citep{pl_bestvina_brady} -- equivalently, if $f(v) \neq f(w)$ holds whenever $\{v,w\}$ forms an edge of $K$. 
Let us fix a PL Morse function $f:K_0 \to \mathbb{R}$; for each real number $c$ we define the \textit{superlevel set} $K^{\geq c}$ as the subcomplex of $K$ spanned by all simplices $\sigma$ whose vertices $v \in \sigma$ satisfy $f(v) \geq c$. 
PL Morse theory aims to explicitly describe how the homology groups of $K^{\geq c}$ evolve as a function of $c$. 

The \textit{upper link} of a vertex $v$ with respect to $f$, denoted $\mathrm{Lk}^+(f;v)$, is defined as the (possibly empty) simplicial subcomplex of $K$ generated by all simplices $\sigma \in K$ such that $\{v\} \cup \sigma$ is a face of $K$, but $v$ is not a face of $\sigma$ and $f(w) \geq f(v)$ for all $w \in \sigma_0$.
It follows that $\mathrm{Lk}^+(f;v)$ has dimension at most $n-1$. 
Given a pair of real numbers $c > d$, note that we automatically have a containment $K^{\geq c} \subset K^{\geq d}$. 
The \textit{link criterion} of PL Morse theory is as follows: if every vertex $v \in K_0$ with $c \geq f(v) \geq d$ has an acyclic $\mathrm{Lk}^+(f;v)$, then the homology groups of $K^{\geq c}$ coincide with those of $K^{\geq d}$. 
Thus, the superlevel set homology can  only change across those $c \in \mathbb{R}$ satisfying $f(v)=c$ for some vertex $v$ whose upper link is \textit{not} acyclic \citep{brehm_kuhnel_1987%
, pl_morse_low_dim%
}. 
These special vertices $v$ are called the \textit{critical points} of $f$, and the corresponding real numbers $c = f(v)$ are the \textit{critical values} of $f$. 
(See Appendix~\ref{appendix:plmorse} for an illustration of how to determine whether or not a vertex is critical based upon its upper link.)

\section{The Morse transform}
\label{sec:morse_transform}

Consider a finite simplicial complex $K \subset \mathbb{R}^n$. 
Given a direction vector $\xi$ lying on the unit sphere $\mathbb{S}^{n-1} \subset \mathbb{R}^n$, let $f_\xi:K_0 \to \mathbb{R}$ be the inner-product map $v \mapsto \langle v,\xi \rangle$. 
Under generic conditions, the map $f_\xi$ is a PL Morse function for which no two critical points have the same critical value. 
Let $N_\xi$ be the number of critical points of $f_\xi$;  
therefore, we may order the critical values of $f_\xi$ as $c^\xi_1 > \cdots > c^\xi_{N_\xi}$ and write $v^\xi_i$ for the unique critical point with value $c^\xi_i \coloneq f_\xi(v^\xi_i)$. 
For brevity, let the $j$-th reduced Betti number of the upper link $\mathrm{Lk}^+(f_\xi;v^\xi_i)$ be denoted by $\widetilde{\beta}(\xi)_j^{i}$. 
It can be shown that $\widetilde{\beta}(\xi)_j$ is zero whenever $j \geq n-1$. 

We define the \textit{Morse transform (MT)} of $K$ (of depth $d \in \mathbb{N}$) along $\xi$ to be a function
\begin{equation*}
\mathscr{M}_{K,d}: \mathbb{S}^{n-1} \to \coprod_{k=0}^{d}\mathrm{Mat}_\mathbb{R}(k,n+1) 
\end{equation*}
that sends a unit vector $\xi \in \mathbb{S}^{n-1}$ to a certain $m \times (n+1)$ matrix of real numbers, where $m \coloneq \mathrm{min}\{d, N_\xi\}$. 
The $i$-th row of this matrix catalogues the relevant Morse data of $f_\xi$ at the $i$-th critical point $v_i^\xi$; explicitly, it is given by
\begin{equation*}
\mathscr{M}_{K,d}(\xi)_{i} \coloneq \left[~c^\xi_i \quad \widetilde{\beta}(\xi)_{-1}^{i} \quad \widetilde{\beta}(\xi)_0^{i} \quad \cdots \quad \widetilde{\beta}(\xi)_{n-2}^{i}~\right].
\end{equation*}
That is to say, the first column of $\mathscr{M}_{K,d}(\xi)$ contains the top $m$ critical values of $f_\xi$ in descending order, whilst the remaining columns, which are all integer-valued, record the reduced Betti numbers of the corresponding upper links. 
We define the \textit{PL Morse index} $\mu$ of a point to be the vector of reduced Betti numbers of its upper link. 
(See Appendix~\ref{appendix:mt_example} for an example of computing the MT for a small simplicial complex.)

We also define a variant of the MT, the \textit{Morse Euler-critical transform (MET)}, similarly to the MT except
each row of the MET records the height and Euler characteristic $\chi(\xi)_{i}$ of the upper link of the $i$-th \textit{Euler-critical} point of $f_\xi$, which is defined to be a point with $\chi(\xi)_{i} \neq 1$. 
The MT contains strictly more information than the MET as the Euler characteristic of a complex can be obtained from its Betti numbers via the Euler-Poincar\'e formula.
(Refer to Appendix~\ref{appendix:met} for more details.) For instance, the MET is blind to critical vertices with Morse index $(0,1,1)$, since their net contribution to the Euler characteristic is zero.

\begin{proposition}
\label{prop:determine}
The Morse transform and the Morse Euler-critical transform at full depth $d = |K_0|$ determine the Euler characteristic transform.
\end{proposition}
For a proof see Appendix~\ref{appendix:transform_comparisons}. 
As a corollary of Proposition~\ref{prop:determine}, at maximal depth both the MT and the MET inherit the injectivity of the ECT on finite simplicial complexes embedded in $\mathbb{R}^n$ \citep{pht_how_many_directions}. The depth-truncated Morse transform used in practice has $d \ll |K_0|$; this trades injectivity in favour of prioritising exterior geometry.

\subsection{Efficient computation}

The Morse transform involves calculating the ranks of homology groups of upper links. This computation has a time complexity of $\mathcal{O}(|\mathrm{Lk}^+|^{3})$ 
when using row reduction or 
$\mathcal{O}(|\mathrm{Lk}^+|^{2.371339})$ 
via fast square matrix-multiplication \citep{roadmap, fast_matrix_exponent}. 
In practice, links of vertices are local subcomplexes of size much smaller than the total number of simplices. 
Moreover, each link homology can be computed in parallel. 
Before assessing the criticality of each vertex along one direction, 
ordering the vertices by height incurs an initial $\mathcal{O}(|K_0| \log |K_0|)$ cost.

It is possible to further improve upon this for simplicial complexes embedded in $\mathbb{R}^n$ for $n \leq 4$, as described below. If the upper link is empty, then 
$\widetilde{\beta}_{i} = \delta_{i,-1}$ by definition.
If the upper link is non-empty, then $\widetilde{\beta}_{-1} = 0$ by definition and:
\begin{itemize}
    \item For $n = 2$, $\widetilde{\beta}_{0}$ can be obtained by counting the number of connected components $\beta_{0} = \widetilde{\beta}_{0} + 1$ with a depth first search (DFS) of the $1$-skeleton of the upper link. 
    Time complexity: 
    $\mathcal{O}(|\mathrm{Lk}^+_1|)$. 
    
    \item For $n = 3$, $\widetilde{\beta}_{0}$ likewise can be calculated by DFS and $\widetilde{\beta}_{1}$ via the Euler characteristic $\chi  = 1 + \widetilde{\beta}_{0} - \widetilde{\beta}_{1}$. 
    Time complexity: 
    $\mathcal{O}(|\mathrm{Lk}^+|)$. 
    
    \item For $n = 4$, $\widetilde{\beta}_{0}$, $\widetilde{\beta}_{1}$ and $\widetilde{\beta}_{2}$ can be computed by the algorithm of \citet{incremental_betti_1995} 
    using a facet-edge data structure.
    Time complexity: 
    $\mathcal{O}(|\mathrm{Lk}^+|)$. 
\end{itemize}
Thus, in each case, for $n \leq 4$, computing the positive link homology at any single vertex incurs a linear cost, which coincides with the asymptotic cost of computing 
the Euler characteristic.
We do not take advantage of the most efficient low-dimensional algorithms for our code implementation of the Morse transform.

\subsection{Vectorisation}

We propose to vectorise the Morse transform by computing summary statistics, inspired by the empirical success of persistence statistics for vectorising persistence diagrams \citep{pd_vectorizations}.
We call our vectorisation \textit{Morse statistics}.
Specifically, consider the Morse transform of a finite simplicial complex $K \subset \mathbb{R}^n$ over a finite-set of directions 
$\xi \in \Xi$.
Let $\mathcal{C}_i$ be the multi-set union of the $i$-th columns of the Morse transform $\mathscr{M}(\xi)^{i}$ for each integer $1 \leq i \leq n + 1$ across all directions, and let $\mathcal{N}$ be the multi-set union of the number of critical points $m_\xi$ for each direction across all directions. 
We compute the $p$-th percentiles of the multi-sets $\mathcal{C}_i$ and $\mathcal{N}$ for all $p$ in $\{0,10,25,40,50,60,75,90,100\}$. 
Then, we concatenate all the percentiles of the multi-sets together to obtain the Morse statistics vector living in $\mathbb{R}^{9(n+2)}$.

This vectorisation is invariant to the arbitrary labelling of directions. 
Furthermore, the length of the vector is independent of the depth of the Morse transform and the number of directions, allowing for classification of sets of heterogenous Morse statistics vectors 
generated from varying numbers of directions. 
A limitation of this vectorisation is that it obfuscates the explicit directional information associated to each critical vertex as it only summarises the distribution of local Morse data. 
Although the MT itself is injective on the space of shapes, the Morse statistics vectorisation is not guaranteed to be injective.

\subsection{Comparison to other topological transforms}

The MT differs from other directional topological transforms like the PHT and ECT that map directions to a collection of persistence diagrams or Euler curves, respectively, instead of matrices of Morse data. 
The MT provides more explicit local information around each critical vertex than the PHT and ECT as the reduced Betti numbers of the upper link allow for a categorisation of PL critical points that is directly analogous to classifying smooth critical points as maxima, minima or saddles. 
In contrast to the the PHT and ECT, the MT naturally focuses on the top $d$ critical vertices closest to the outer surface of a shape, which is the most consequential region for tasks where surface geometry is pertinent, like protein-ligand binding.
For low dimensions ($\leq 4$), calculating the Morse transform along one direction is asymptotically equal in worst-case time complexity to the ECT $\mathcal{O}(|K_0| \log |K_0|)$ and faster than the PHT $\mathcal{O}(|K|^{2.371339})$.
(See Appendix~\ref{appendix:transform_comparisons} for a more comprehensive comparison of the Morse transform with other topological transforms.)

\section{Methodology}
\label{sec:methodology}

\subsection{Molecules as simplicial complexes}

There are several discrete models for representing molecular data; the molecular graph is the most iconic, but it lacks intrinsic 3D structure. Among the most viscerally geometric is the space-filling model or \textit{union of balls}, where one constructs 3D balls (of van der Waals radius) around atom centres \citep{shapes_of_macromolecules}. This representation is visually appealing but rather unwieldy from a computational perspective -- checking whether or not a point in this union lies on the boundary is a cumbersome task.  
Therefore, it is common to represent a given molecule as a simplicial complex \citeprefix{Ch.~3.1 of}{spanier} whose vertices correspond to atom centres. 

For the purposes of building such a simplicial model, suppose we have access to a finite subset $P$ of Euclidean space $\mathbb{R}^3$ (known as a point cloud) consisting of atom-centres of a given molecule and a function $w:P \to \mathbb{R}$ that sends each atom $p \in P$ to the corresponding van der Waals radius $w_p$. The weighted distance of a point $x \in \mathbb{R}^3$ to $p$ is the real number $d_w(x,p) \coloneq \|p-x\|^2-w_p^2$, where $\|\bullet\|$ denotes the standard Euclidean norm. The \textit{weighted Voronoi cell} of $p \in P$ is the subset $V(p) \subset \mathbb{R}^3$ given by
\[
V_w(p) \coloneq \left\{x \in \mathbb{R}^3 \mid d_w(x,p) \leq d_w(x,q) 
\: \forall
q \in P \setminus\!\{p\}\right\}.
\]
Explicitly, this consists of all points in $\mathbb{R}^3$ which admit $p$ as a nearest neighbour (in $P$, with respect to $d_w$). Each $V_w(p)$ is a closed convex set and the collection of $\left\{V_w(p) \mid p \in P\right\}$ forms a regular cell decomposition of $\mathbb{R}^3$ (consult \citet{aurenhammer} for details).

\begin{figure*}[ht!]%
\centering
\includegraphics[width=1.0\linewidth]{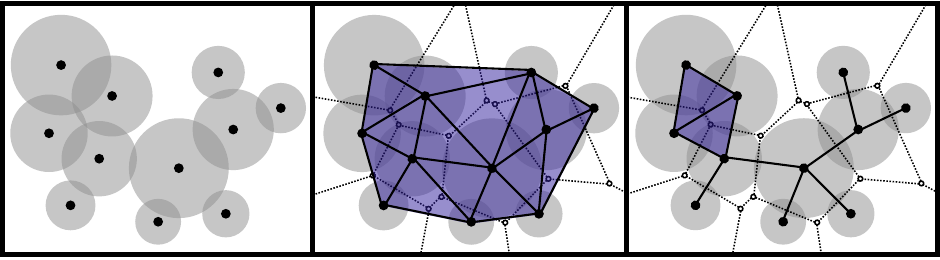}
\caption{Generation of a pruned Delaunay triangulation. \textbf{Left panel:} we begin with a point cloud along with a set of weights, which encode the radius of each weighted point. This weighted point cloud can be represented as a union of balls (grey circles).
\textbf{Centre panel:} the space is partitioned into weighted Voronoi cells (dotted lines) and the dual weighted Delaunay triangulation is constructed (solid points, lines and blue triangles). 
\textbf{Right panel:} we prune the triangulation of any edges that connect two weighted points whose associated balls do not overlap. 
}
\label{fig:molecular_complex_generation}
\end{figure*}

When $P$ is in general position, the dual Voronoi cellulation forms a simplicial complex, which is called the  \textit{weighted Delaunay triangulation} \citep{edelsbrunner} of $P$ and denoted $D_w(P) \subset \mathbb{R}^3$. More precisely,
the vertex set of $D_w(P)$ is $P$ and there is a $k$-simplex for $k \in \{1,2,3\}$ spanning $\{p_0,p_1,\ldots,p_k\}$ whenever the corresponding weighted Voronoi cells have nonempty intersection, i.e. when $\bigcap_{j=0}^k V_w(p_j) \neq \varnothing$. It is possible for simplices in $D_w(P)$ to contain edges which are too long to accurately reflect the underlying molecular geometry. Hence, we remove every simplex from $D_w(P)$ that contains an edge $\{p,q\}$ whose length $\|p-q\|$ exceeds the sum $w_p+w_q$ of the associated van der Waals radii. The resulting simplicial subcomplex $K_w(P) \subset D_w(P)$, which we call the \textit{pruned Delaunay triangulation} of $P$, serves as a convenient geometric representation of each given molecule (Fig.~\ref{fig:molecular_complex_generation}).

\subsection{The Morse molecular feature vector} 
\label{methods:feature_vector}

Let $P' \subset \mathbb{R}^3$ be the collection of atom-centres of a candidate molecule. 
First, we construct a set $\Xi$ of $100$ directions in the unit sphere $\mathbb{S}^2$ as follows: $32$ directions correspond to vertices of the (origin-centred) pentakis dodecahedron; and the remaining $68$ directions are chosen at random by uniformly sampling $\mathbb{S}^2$. 
Our feature vector is constructed in five steps (MORSE):
\begin{enumerate}
    \item {\bfseries Modify the point cloud:} we centre $P'$ about the origin and then apply a rotation so that the co\"{o}rdinate axes coincide with the principal components (further details in Appendices~\ref{appendix:rot_inv} 
    and \ref{appendix:pca_analysis}). 
    Let us call the resulting point cloud $P \subset \mathbb{R}^3$.
    \item {\bfseries Obtain the triangulation:} we construct the pruned Delaunay triangulation $K \coloneq K_w(P)$, where $w:P \to \mathbb{R}$ maps  each vertex to the van der Waals radius of the corresponding atom.
    \item {\bfseries Realise the Morse transform:} we compute the Morse transform $\mathscr{M}_{K,d}$ of depth $d = 20$ along the directions $\Xi$. Let $m_\xi \leq d$ be the total number of critical points found along the direction $\xi \in \Xi$ up to depth $d$. Since $K \subset \mathbb{R}^3$, the Morse transform maps each direction $\xi$ to an $m_\xi \times 4$ matrix $\mathscr{M}(\xi)$. We denote the $i$-th column of $\mathscr{M}(\xi)$ by $\mathscr{M}^{i}(\xi)$. 
    \item {\bfseries Supplement with non-shape
    data:} for each atom in the molecule we insert the contribution of the atom to the partial charge, molar refractivity and lipophilicity into three multi-sets $\mathcal{Q}$, $\mathcal{R}$ and $\mathcal{L}$, respectively (see Appendix~\ref{appendix:virtual}).
    \item {\bfseries Encapsulate in one vector:} let $\mathcal{C}_i$ be the multi-set union of the $i$-th columns $\mathscr{M}(\xi)^{i}$ for each integer $1 \leq i \leq 4$ across all directions, and let $\mathcal{N}$ be the multi-set union of the number of critical points $m_\xi$ across all directions. 
    We compute the $p$-th percentiles of the multi-sets $\mathcal{C}_i$, $\mathcal{N}$, $\mathcal{Q}$, $\mathcal{R}$ and $\mathcal{L}$ for all $p$ in $\{0,10,25,40,50,60,75,90,100\}$. Finally, we concatenate all the percentiles of the multi-sets together to obtain a single vector of length $72$. 
\end{enumerate} 
This constitutes the chemistry-supplemented \textit{Morse feature vector} that we assign to the molecule represented by $P' \subset \mathbb{R}^3$. 
(In essence, this is the Morse statistics vectorisation concatenated with chemical information.)

\subsection{Classifier}
\label{methods:classifier}

\begin{figure}[ht!]
\centering
\includegraphics[width=1.0\linewidth]{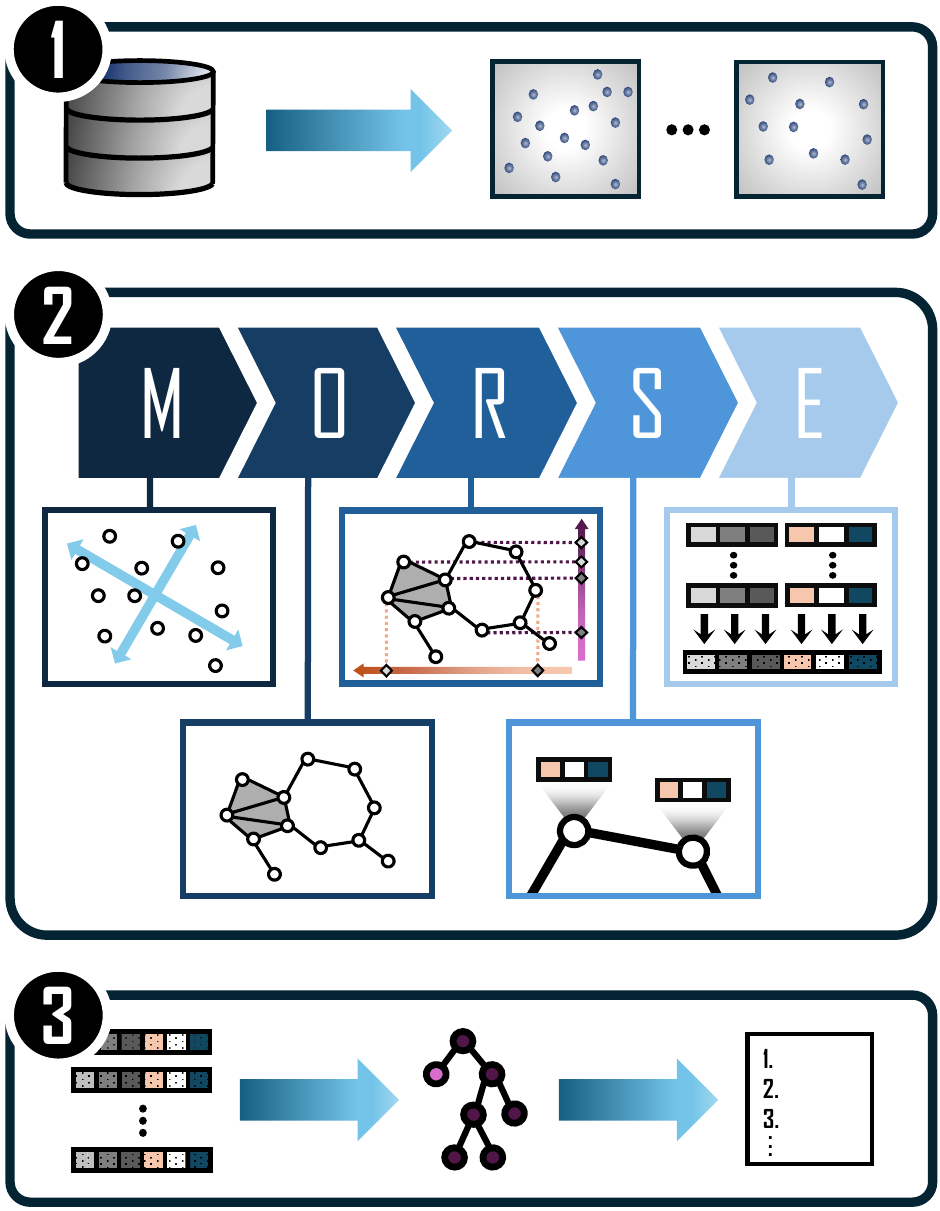}
\caption{A diagrammatic overview of our LBVS pipeline. 
(1) \textit{Data preparation.} We generate molecular point clouds from the datasets DUD-E and MUV (see Section~\ref{methods:dataset_preparation}). 
(2) \textit{Feature generation.} We compute our topological feature vector in five steps 
(\textbf{MORSE}).  
See Section~\ref{methods:feature_vector} for more details. 
(3) \textit{Classification.} We train a gradient-boosted decision tree on Morse feature vectors to perform a binary classification task to rank molecules on how likely they are to bind to specific protein targets (see Section~\ref{methods:classifier}).}
\label{fig:overview}
\end{figure}

We train a Light Gradient Boosting Machine \citeprefix{LGBM;}{lightgbm} for binary classification on Morse features (and comparison descriptors from the literature). We tune and evaluate the classifier using $5$-fold cross-validation (for details, see Appendices~\ref{appendix:tuning} and \ref{appendix:metrics}). The final evaluation metric scores are the mean scores across all the $5$-folds. 
A diagrammatic summary of our entire pipeline is displayed in Fig.~\ref{fig:overview}.

\subsection{Visualisation}

As a part of our code implementation, we provide a way to interactively visualise 3D molecular simplicial complexes and annotate the atoms according to their PL Morse indices for a particular height function. 
(See Appendix~\ref{appendix:visualisation} for examples of visualisations.)

\section{Experiments}
\label{sec:experiments}

\subsection{Datasets}

We evaluate the Morse transform-based shape summary on two widely used benchmark datasets for virtual screening: the Directory of Useful Decoys, Enhanced \citeprefix{DUD-E;}{dude} and the Maximum Unbiased Validation \citeprefix{MUV;}{muv} datasets. DUD-E consists of 102 protein targets, each of which is assigned active ligands that have been experimentally confirmed to have high binding affinity and property-matched decoy molecules.
Decoy molecules resemble the ligands physically but are dissimilar in 2D chemical fingerprint space to minimise the chance of binding and are presumed to be inactive.
The DUD-E Diverse (D8) subset is a subset of DUD-E consisting of 8 targets that are representative of the diverse protein categories in DUD-E. In contrast, MUV consists of 17 protein targets, each of which is assigned active and decoy molecules that are both verified by experimental PubChem bioassay data. MUV was designed to not be affected by artificial enrichment or analogue bias by ensuring that actives are close to decoys in simple chemical descriptor space.

\subsection{Dataset preparation} 
\label{methods:dataset_preparation}

We use a pipeline based on \textsc{rdk}{\small it} to curate and prepare DUD-E and MUV molecules for descriptor generation. 
Firstly, we remove duplicate molecules by comparing their canonical Simplified Molecular Input Line Entry System \citeprefix{SMILES;}{smiles} representations. 
Then we standardise the molecules to ensure a consistent representation.
The datasets contain molecules with varying protonation states; therefore, a ‘wash’ step is included so that each molecule is in a standard protonation state. 
We generate multiple conformations using \textsc{rdk}{\small it} \citep{conformer}, and after energy-minimisation the conformer with the lowest calculated force-field energy is retained and used for subsequent analysis.

\subsection{Classification task}

We test the performance of a LGBM classifier trained on various molecular descriptors on a series of binary classification tasks to distinguish actives from decoys for each target in D8 and MUV.
(The tuning and evaluation are the same as in Section~\ref{methods:classifier}.)

\subsection{Comparison topological descriptors}

We investigate the performance of descriptors derived from directional topological transforms.
We evaluate two versions of our features: $45$-dimensional plain Morse features at depth 20 and 32 directions without the chemical percentiles (MT-P); 
and $27$-dimensional plain Morse Euler-critical features at depth 20 and 32 directions without the chemical percentiles (MET-P). 
We compare against the following ECT-based descriptors: $5\,973$-dimensional flattened ECT sampled at 181 thresholds and 32 directions (ECT-F); and $10$-dimensional concatenated percentiles of the same Euler characteristic curves using the same percentiles of the Morse feature vector (ECT-P).
(See Appendix~\ref{appendix:ect_vec} for more details about our ECT vectorisations.) 
We do not compare against any PHT-based descriptors as there are no mature implementations available and there is no canonical choice from the multitude of persistence diagram vectorisations \citep{pd_vectorizations} to create 
an agglomerate PHT vectorisation.

\subsection{LBVS method comparison criteria}

There is a smorgasbord of virtual screening methods in the literature and an almost commensurate number of testing methodologies. Therefore, to ensure a fair comparison we only evaluate our features against shape-based molecular descriptors that satisfy the following criteria:
\begin{itemize}
\item \textbf{LBVS descriptor:} we exclude descriptors that use any explicit information about the protein binding site, such as structure-based virtual screening methods.

\item \textbf{Non-superpositional descriptor:} we exclude superpositional descriptors that can only rank molecules using a similarity metric as these generally perform worse than machine learning classifiers trained on non\-/superpositional descriptors 
(see Appendix~\ref{appendix:similarity}).

\item \textbf{Similar testing procedure:} we only directly compare against methods that can be cross-validated and do not use only a single data split. (Various deep learning methods are not suitable for comparison due to incompatible methodological differences, see Section~\ref{subsec:deep_learning} for case studies.)

\item \textbf{Code shared online:} we only compare against methods that have an implementation freely available online.
\end{itemize}

\subsection{Comparison LBVS descriptors}

Traditionally, LBVS methods involve generating hand-crafted molecular \textit{descriptors} (features), which aim to numerically capture the relevant chemical and physical properties of ligands \citep{descriptor_book_todeschini}. 
Most molecular descriptors can be categorised as 0D, 1D, 2D or 3D descriptors (shape-based), depending upon the kind of molecular information they encode \citep{descriptors_survey, descriptors_taxonomy_ml}. 

We evaluate two versions of our features: 45-dimensional plain Morse features at depth 20 and 100 directions without the chemical percentiles (M); and the 72-dimensional chemistry-supplemented Morse features at depth 20 and 100 directions (M+C). We also test a 27-dimensional baseline descriptor consisting of nine percentiles of each 3D ordinate of the atoms of a randomly rotated molecule (3DP). 
We compare against the following shape-based LBVS descriptors: the unweighted subset of Weighted Holistic Invariant Molecular descriptors \citeprefix{WHIMu;}{whim}; Ultrafast Shape Recognition \citeprefix{USR;}{usr}; Riemannian Geometry for Molecular Surface Approximation \citeprefix{RGMolSA;}{rgmolsa}; and K\"ahler Quantisation for Molecular Surface Approximation \citeprefix{KQMolSA;}{kqmolsa}. Additionally, we compare against the following hybrid shape- and chemistry-based descriptors: the full set of Weighted Holistic Invariant Molecular descriptors (WHIM); Ultrafast Shape Recognition with CREDO Atom Types \citeprefix{USRCAT;}{usrcat}; and Weighted Holistic Atom Localisation and Entity Shape descriptors \citeprefix{WHALES;}{whales2018, whales2021}.

\subsection{Results}

\begin{table}[!ht]
\centering
\caption{The mean AUROC per D8 and MUV target and the overall mean AUROC of various topological transform-based descriptors. For each target the descriptor(s) with the highest AUROC is displayed in bold. SD is the sample standard deviation of the mixture distribution of the target performances. 
}
\label{tab:dude_muv_topo_auroc}
\footnotesize
\def\angle{90}
%
\def\bigspace{8pt}
\newcolumntype{L}{l<{\hspace{\bigspace}}}
\newcolumntype{C}{c<{\hspace{\bigspace}}}
\begin{tabular}{lLcccc}
\toprule 
& & \multicolumn{4}{C}{Topological} \\
\cmidrule(lr){3-6} 
Dataset & Target 
& \rotatebox{\angle}{MT-P} 
& \rotatebox{\angle}{MET-P} & \rotatebox{\angle}{ECT-F} & \rotatebox{\angle}{ECT-P} \\
\midrule
\multirow{8}{*}{D8} & AKT1 & $\bf 0.88$ & $0.85$ & $0.81$ & $0.69$\\
& AMPC & $\bf 0.82$ & $0.79$ & $0.69$ & $0.69$\\
& CP3A4 & $\bf 0.66$ & $0.64$ & $0.59$ & $0.54$\\
& CXCR4 & $0.86$ & $\bf 0.88$ & $0.78$ & $0.61$\\
& GCR & $0.86$ & $0.85$ & $\bf 0.87$ & $0.78$\\
& HIVPR & $\bf 0.93$ & $0.92$ & $0.90$ & $0.81$\\
& HIVRT & $\bf 0.81$ & $\bf 0.81$ & $\bf 0.81$ & $0.74$\\
& KIF11 & $\bf 0.90$ & $0.89$ & $0.89$ & $0.74$\\
\cmidrule(lr){2-6}
& mean & $\bf 0.84$ & $0.83$ & $0.79$ & $0.70$\\
& SD & $0.08$ & $0.09$ & $0.11$ & $0.09$\\
\midrule
\multirow{17}{*}{MUV} & 466 & $0.54$ & $0.57$ & $\bf 0.58$ & $0.56$\\
& 548 & $\bf 0.71$ & $0.69$ & $0.58$ & $0.67$\\
& 600 & $0.61$ & $0.51$ & $0.55$ & $\bf 0.65$\\
& 644 & $0.69$ & $\bf 0.70$ & $0.60$ & $0.66$\\
& 652 & $\bf 0.64$ & $\bf 0.64$ & $0.59$ & $0.49$\\
& 689 & $\bf 0.64$ & $0.60$ & $0.47$ & $0.60$\\
& 692 & $\bf 0.69$ & $0.64$ & $0.67$ & $0.66$\\
& 712 & $\bf 0.72$ & $0.68$ & $0.62$ & $0.63$\\
& 713 & $0.67$ & $\bf 0.68$ & $0.60$ & $0.63$\\
& 733 & $0.63$ & $\bf 0.66$ & $0.47$ & $0.58$\\
& 737 & $\bf 0.73$ & $0.71$ & $0.58$ & $0.68$\\
& 810 & $\bf 0.62$ & $\bf 0.62$ & $0.51$ & $0.50$\\
& 832 & $0.65$ & $\bf 0.67$ & $0.55$ & $0.44$\\
& 846 & $0.77$ & $\bf 0.78$ & $0.65$ & $0.72$\\
& 852 & $0.68$ & $0.65$ & $\bf 0.71$ & $\bf 0.71$\\
& 858 & $0.55$ & $0.52$ & $\bf 0.61$ & $0.59$\\
& 859 & $0.57$ & $0.56$ & $0.52$ & $\bf 0.62$\\
\cmidrule(lr){2-6}
& mean & $\bf 0.65$ & $0.64$ & $0.58$ & $0.61$\\
& SD & $0.11$ & $0.11$ & $0.12$ & $0.12$\\
\bottomrule
\bottomrule
\end{tabular}
\end{table}

\begin{table*}[!ht]
\centering
\caption{The mean AUROC per D8 and MUV target and the overall mean AUROC of various LBVS descriptors. For each target the descriptor(s) with the highest AUROC is displayed in bold and the highest shape-based descriptor is underlined. SD is the sample standard deviation of the mixture distribution of the target performances. 
}
\label{tab:dude_muv_auroc}
\footnotesize
\def\angle{90}
%
\def\bigspace{8pt}
\newcolumntype{L}{l<{\hspace{\bigspace}}}
\newcolumntype{C}{c<{\hspace{\bigspace}}}
\begin{tabular}{lLcccccCcccc}
\toprule 
& & \multicolumn{6}{C}{Shape} & \multicolumn{4}{c}{Shape \& Chemistry}\\
\cmidrule(lr{\bigspace}){3-8} \cmidrule(lr){9-12}
Dataset & Target & \rotatebox{\angle}{M} & \rotatebox{\angle}{3DP} & \rotatebox{\angle}{USR} & \rotatebox{\angle}{WHIMu} & \rotatebox{\angle}{RGMolSA} & \rotatebox{\angle}{KQMolSA} & \rotatebox{\angle}{M+C} & \rotatebox{\angle}{USRCAT} & \rotatebox{\angle}{WHIM} & \rotatebox{\angle}{WHALES}\\
\midrule
\multirow{8}{*}{D8} & AKT1 & $\underline{0.86}$ & $0.65$ & $0.77$ & $0.82$ & $0.83$ & $0.75$ & $\bf 0.99$ & $0.98$ & $0.92$ & $0.93$\\
& AMPC & $\underline{0.82}$ & $0.56$ & $0.74$ & $0.79$ & $0.66$ & $0.57$ & $\bf 0.94$ & $0.88$ & $0.84$ & $0.93$\\
& CP3A4 & $0.65$ & $0.52$ & $0.53$ & $0.60$ & $\underline{0.68}$ & $0.62$ & $\bf 0.92$ & $0.79$ & $0.72$ & $0.79$\\
& CXCR4 & $\underline{0.87}$ & $0.61$ & $0.81$ & $0.78$ & $0.84$ & $0.70$& $\bf 0.99$ & $0.93$ & $0.87$ & $0.90$\\
& GCR & $\underline{0.88}$ & $0.65$ & $0.83$ & $0.87$ & $0.84$ & $0.67$ & $\bf 0.99$ & $0.94$ & $0.95$ & $0.90$\\
& HIVPR & $\underline{0.93}$ & $0.71$ & $0.83$ & $0.90$ & $0.92$ & $0.83$ & $\bf 0.99$ & $0.96$ & $0.96$ & $0.91$\\
& HIVRT & $\underline{0.82}$ & $0.63$ & $0.75$ & $0.78$ & $0.77$ & $0.58$ & $\bf 0.97$ & $0.90$ & $0.85$ & $0.82$\\
& KIF11 & $0.89$ & $0.67$ & $0.87$ & $\underline{0.91}$ & $0.86$ & $0.77$ & $\bf 0.98$ & $\bf 0.98$ & $0.96$ & $0.94$\\
\cmidrule(lr){2-12}
& mean & $\underline{0.84}$ & $0.62$ & $0.76$ & $0.81$ & $0.80$ & $0.69$ & $\bf 0.97$ & $0.92$ & $0.88$ & $0.89$\\
& SD & $0.09$ & $0.08$ & $0.11$ & $0.11$ & $0.10$ & $0.10$ & $0.03$ & $0.07$ & $0.09$ & $0.06$\\
\midrule
\multirow{17}{*}{MUV} & 466 & $\underline{0.59}$ & $0.54$ & $0.56$ & $0.56$ & $0.47$ & $0.46$ & $0.61$ & $0.64$ & $0.61$ & $\bf 0.69$\\
& 548 & $\underline{0.73}$ & $0.57$ & $0.62$ & $0.65$ & $0.65$ & $0.65$ & $\bf 0.83$ & $0.74$ & $0.69$ & $0.64$\\
& 600 & $0.52$ & $0.59$ & $0.58$ & $\underline{0.62}$ & $0.56$ & $0.55$ & $\bf 0.64$ & $0.59$ & $\bf 0.64$ & $0.53$\\
& 644 & $\underline{0.72}$ & $0.57$ & $0.59$ & $0.70$ & $0.63$ & $\underline{0.72}$ & $ 0.79$ & $\bf 0.83$ & $0.75$ & $0.65$\\
& 652 & $0.62$ & $0.57$ & $0.57$ & $0.43$ & $\underline{0.63}$ & $0.40$ & $\bf 0.70$ & $0.57$ & $0.60$ & $0.61$\\
& 689 & $\underline{0.65}$ & $0.62$ & $0.46$ & $0.45$ & $0.57$ & $0.51$ & $\bf 0.74$ & $0.57$ & $0.44$ & $0.62$\\
& 692 & $0.68$ & $0.49$ & $0.62$ & $\bf \underline{0.74}$ & $0.60$ & $0.59$ & $0.65$ & $0.59$ & $0.72$ & $0.65$\\
& 712 & $\underline{0.69}$ & $0.51$ & $0.61$ & $0.62$ & $0.55$ & $0.48$ & $\bf 0.75$ & $0.63$ & $0.71$ & $0.69$\\
& 713 & $\underline{0.69}$ & $0.66$ & $0.65$ & $0.57$ & $0.55$ & $0.49$ & $\bf 0.73$ & $0.70$ & $0.72$ & $0.68$\\
& 733 & $\underline{0.66}$ & $0.45$ & $0.44$ & $0.53$ & $0.57$ & $0.43$ & $\bf 0.67$ & $0.45$ & $0.59$ & $0.59$\\
& 737 & $\underline{0.75}$ & $0.50$ & $0.63$ & $0.65$ & $0.59$ & $0.57$& $\bf 0.81$ & $0.65$ & $0.67$ & $0.74$\\
& 810 & $0.54$ & $0.43$ & $0.46$ & $\underline{0.62}$ & $0.56$ & $0.58$ & $\bf 0.73$ & $0.48$ & $0.61$ & $0.61$\\
& 832 & $\underline{0.68}$ & $0.43$ & $0.52$ & $0.59$ & $0.48$ & $0.49$ & $\bf 0.81$ & $\bf 0.81$ & $0.53$ & $\bf 0.81$\\
& 846 & $\underline{0.79}$ & $0.47$ & $0.45$ & $0.73$ & $0.53$ & $0.64$ & $\bf 0.90$ & $0.70$ & $0.81$ & $0.80$\\
& 852 & $\underline{0.73}$ & $0.55$ & $0.67$ & $0.70$ & $0.57$ & $0.53$ & $\bf 0.81$ & $0.73$ & $0.67$ & $0.70$\\
& 858 & $0.58$ & $0.49$ & $0.57$ & $\bf \underline{0.73}$ & $0.63$ & $0.48$ & $0.72$ & $0.60$ & $0.69$ & $0.65$\\
& 859 & $\bf \underline{0.64}$ & $0.52$ & $0.53$ & $0.54$ & $0.58$ & $0.40$ & $0.61$ & $0.41$ & $0.50$ & $0.56$\\
\cmidrule(lr){2-12}
& mean & $\underline{0.66}$ & $0.53$ & $0.56$ & $0.61$ & $0.57$ & $0.53$ & $\bf 0.74$ & $0.63$ & $0.64$ & $0.66$\\
& SD & $0.12$ & $0.10$ & $0.12$ & $0.14$ & $0.11$ & $0.12$ & $0.12$ & $0.15$ & $0.14$ & $0.12$\\
\bottomrule
\bottomrule
\end{tabular}
\end{table*}
\subsubsection*{Topological transforms}

Table~\ref{tab:dude_muv_topo_auroc} displays the mean and per target performance of a variety of descriptors generated from vectorisations of topological transforms.
On the D8 subset, MT-P features score the best with a mean of $0.84 \pm 0.08$ (highest or joint-highest score across $6/8$ individual targets) and MET-P are second-best with a mean of $0.83 \pm 0.09$.
On MUV, similarly MT-P features score the best with a mean of $0.65 \pm 0.11$ (highest or joint-highest score across $7/17$ targets) and MET-P features are the second-best with a mean of $0.64 \pm 0.11$.
On both datasets, the ECT-based features score worse than the Morse transform-based features, though the difference is dataset dependent.

\subsubsection*{Specialised LBVS methods}

Table~\ref{tab:dude_muv_auroc} displays the mean and per target performance of a variety of different LBVS descriptors generated from both datasets. 
On both D8 and MUV, Morse features (M) achieve the highest mean performance among the shape-based descriptors. Similarly, the chemistry\-/supplemented Morse features (M+C) are the best performing among all the evaluated descriptors. For some targets plain Morse features score even better than USRCAT, WHIM and WHALES despite containing no explicit chemical information.
Without exception, all descriptors perform on average worse on the MUV dataset than the D8 subset; this confirms that MUV is a more challenging dataset.

Chemistry\-/supplemented Morse features score substantially higher than plain Morse features, demonstrating the importance of chemical information for accurately identifying active ligands. 
Indeed, this trend is also seen for other descriptors where the chemically-enriched versions (USRCAT, WHIM) on average outperform their shape-based counterparts (USR, WHIMu).
Nevertheless, for specific targets in MUV (692, 858 and 859) the shape-based descriptors achieve the best results, which implies that the relative importance of geometric and chemical information is target-dependent.

\begin{figure*}[!ht]
    \centering
    \begin{subfigure}{0.42\linewidth}
        \centering
        \includegraphics[width=\linewidth]{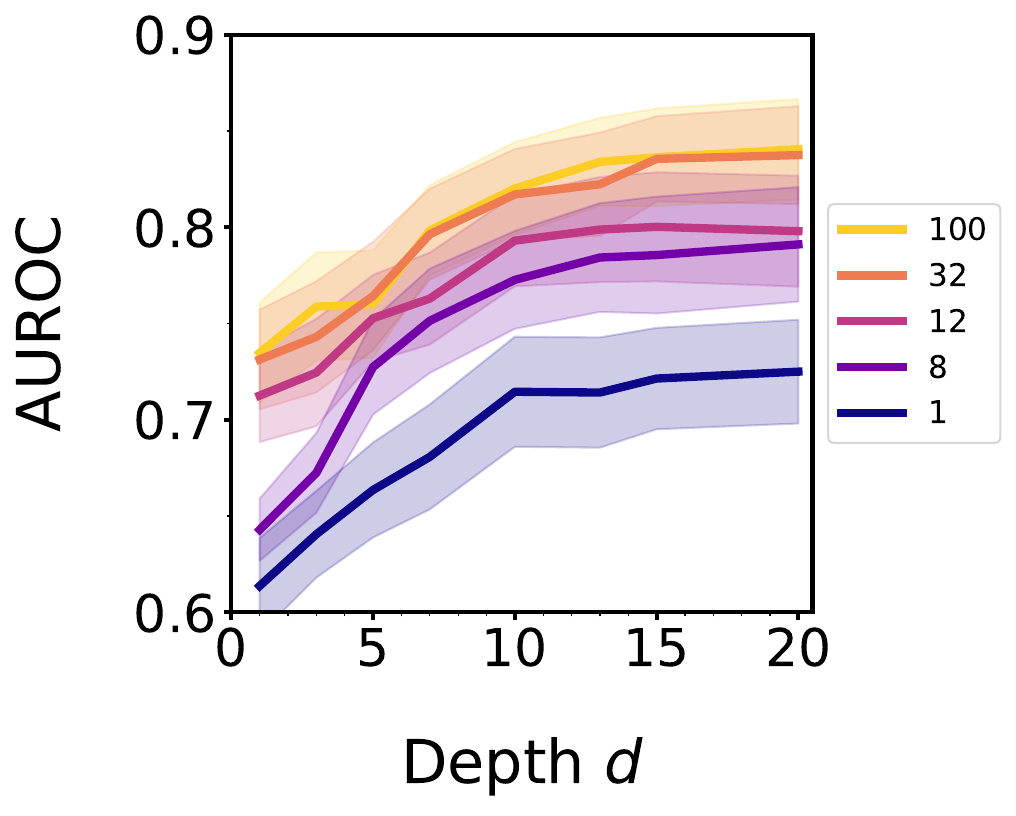}
        \caption{}
        \label{subfig:dude_rocauc_directions_plot}
    \end{subfigure}
    \hspace{1.4cm}
    \begin{subfigure}{0.45\linewidth}
        \centering
        \includegraphics[width=\linewidth]{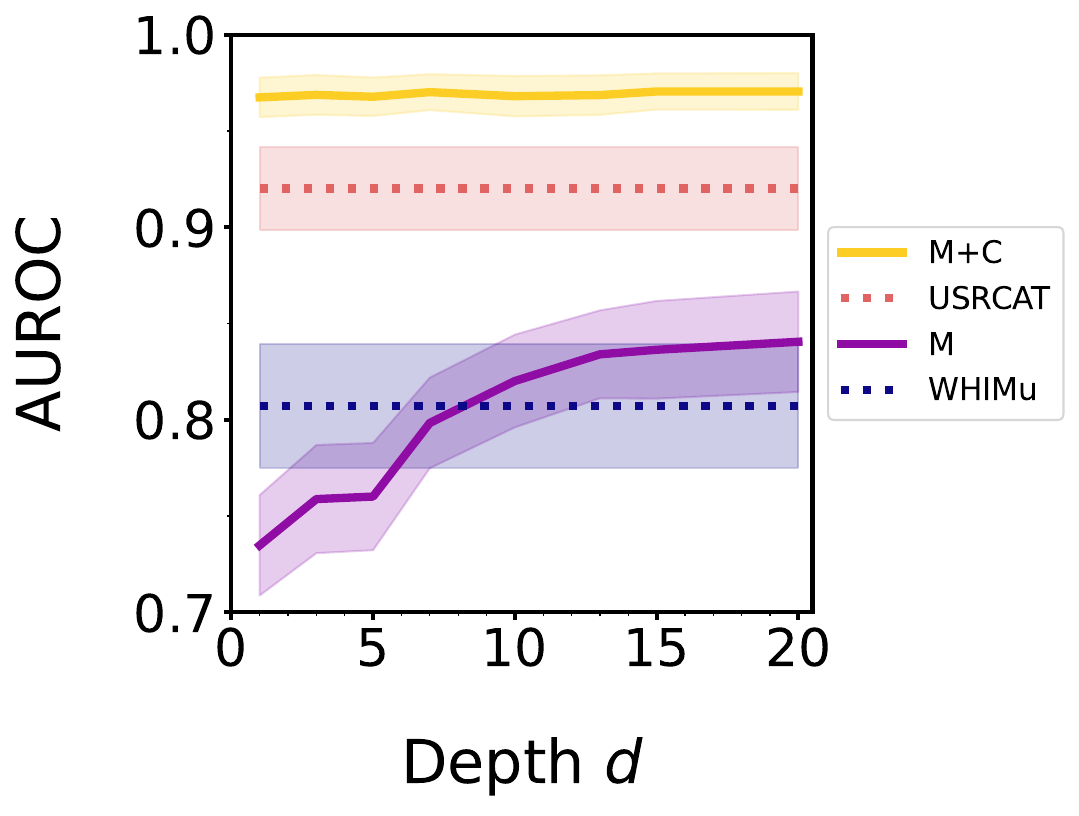}
        \caption{}
        \label{subfig:dude_external_comparison_line_plot}
    \end{subfigure}
    \caption{Panel~\subref{subfig:dude_rocauc_directions_plot}: the mean AUROC score against depth of our LGBM classifier trained on Morse feature vectors computed using 100 directions (100), 32 pentakis dodecahedral directions (32), 12 icosahedral directions (12), 8 cubic directions (8) and 1 direction (1) for the D8 subset. 
    Panel~\subref{subfig:dude_external_comparison_line_plot}: the mean AUROC score against depth of the LGBM classifier trained on chemistry-supplemented Morse feature vectors computed using 100 directions (M+C) and Morse feature vectors computed using 100 directions (M) for the D8 subset. 
    The best-performing external shape- and chemistry-based descriptor USRCAT and the best performing external shape-based descriptor WHIMu are plotted with dotted lines. 
    Error bars are 95\% confidence intervals in both panels.
    }
    \label{fig:dude_depth_ablation}
\end{figure*}

\subsection{Ablation studies}
\label{subsec:robustness}

We evaluate the sensitivity of our Morse LBVS framework through a series of ablation studies of the Morse feature vector. Generating the Morse feature of a molecule involves specifying a set of query directions and the depth of the Morse transform. 
We retrain our classifier on plain Morse features generated from a single direction, 8 cubic directions, 12 icosahedral directions, 32 pentakis dodecahedral directions and 
100 directions, comprising of 68 directions uniformly sampled from the 2-sphere and the 32 pentakis dodecahedral ones, at various depths in $\{1, 3, 5, 7, 10, 13, 15, 20\}$.

We find that increasing both the depth and the number of directions improves the generalisation error with diminishing returns. 
On D8 and MUV, the performance plateaus around a depth of 10 and 32 directions (Fig.~\ref{subfig:dude_rocauc_directions_plot} and Fig.~\ref{fig:muv_auroc_directions_plot}).
The performance is also not substantially affected by the particular choice of 32 directions (see Appendix~\ref{appendix:direction_robustness}).
The lack of any prominent local peaks in the performance before the plateau indicates that Morse features are well-behaved with respect to the depth and number of directions. 
Reassuringly, this means that these parameters do not have to be optimised or tuned to find a local maximum in performance 
-- just selecting a reasonably high value should be sufficient to extract most of the possible performance. 
This direction behaviour is expected as probing a molecule from more directions increases the chance of identifying salient geometric features, and after sufficiently many directions the sample percentiles of the components of the Morse feature should tend to the population percentiles of the Morse feature for an infinite number of directions. 
These results empirically align with theoretical work on topological transforms that proves that only a finite number of directions is required to completely determine a PL shape \citep{pht_how_many_directions}.

The performance of chemistry-supplemented Morse features is relatively unaffected by the depth unlike plain Morse features. 
For all depths, chemistry-supplemented Morse features dominate the performance of the second-best descriptor whilst plain Morse features surpass the performance of the second-best shape-based descriptor at a depth of around 10 (Fig.~\ref{subfig:dude_external_comparison_line_plot} and Fig.~\ref{fig:muv_external_comparison_line_plot}).

Our plain Morse method is robust to positional noise as adding mean-zero Gaussian perturbations to the molecular point clouds does not substantially affect the performance of the classifier until the standard deviation of the noise is large (see Appendix~\ref{appendix:noise}). 
Moreover, modifying the Morse transform to record the Morse data of both the critical \textit{and} non-critical vertices or randomly selecting vertices degrades the performance of the feature (see Fig.~\ref{fig:dude_diverse_auroc_alternate_line} and \ref{fig:muv_auroc_alternate_line}), which empirically justifies the Morse-theoretic approach of focusing on critical vertices.

\subsection{LBVS deep learning model comparison}
\label{subsec:deep_learning}

Recent approaches in virtual \linebreak screening have been dominated by sophisticated deep learning methods that have demonstrated impressive performance in computational benchmarks \citep{wu2024review}. 
Among the best performing of these on DUD-E is DeepScore \citep{deepscore}, which reports a mean AUROC of $0.98$ on all $102$ targets using $5$-fold cross\-/validation and $0.96 \pm 0.03$ when restricted to the D8 subset.
This is only slightly lower than our M+C's $0.97 \pm 0.03$. 

The best performing deep learning models on MUV are Deep-CBN \citep{deepcbn} and TrimNet \citep{trimnet}, which report mean AUROCs of $0.998$ and $0.85$, respectively. After rerunning Deep-CBN with $5$-fold cross-validation we find a mean AUROC of $0.69 \pm 0.11$ for its phase 1 model and a mean AUROC of $0.55 \pm 0.03$ for its final phase 3 model across all MUV targets, both of which are lower scoring than M+C. We are able to replicate TrimNet's mean AUROC of $0.85$ only when using the method's default random seed of $68$, which determines the single test-validation-train data split. The mean AUROC of TrimNet across $10$ random seeds between $2021-2030$ drops to $0.75 \pm 0.03$, demonstrating the sensitivity of the method to the initial data split and the need for cross-validation.\footnote{For comparison, our M+C achieves $0.85 \pm 0.07$ when restricted to the best split from cross-validation for each target.}

These deep-learning results are not part of our primary descriptor comparison because the underlying models use different input modalities, training procedures and (most crucially) validation protocols. Nevertheless, they provide useful context for the scales of the observed AUROC values. Under more comparable validation procedures, the performance of Deep-CBN decreases substantially, whilst TrimNet displays high sensitivity to the choice of random data split. These observations reinforce the need for cross-validation, and show that the proposed M+C descriptor is competitive with substantially more complex learned representations on the datasets considered here.

\section{Conclusion}
\label{sec:conclusion}

We introduced the Morse transform to analyse discrete shapes by characterising their critical points under multiple directional height-functions.
Our work demonstrates the effectiveness of the Morse transform in providing a rich shape representation and empirically showcases the power of the Morse-theoretic approach of concentrating on the local critical point information without needing to consider the whole superlevel set structure, unlike other directional topological transforms. 
The Morse transform has several distinguishing properties:
(i) explicit locality by focusing on critical points; 
(ii) flexible global-to-local dimensional reduction by varying the depth, which prioritises exterior geometry; 
(iii) it employs a finer topological invariant than the Euler characteristic whilst being efficiently computable up to 4D; and
(iv) a compact vectorisation, independent of the labelling of directions.
Our Morse transform-based LBVS method on average outperformed methods derived from the Euler characteristic transform and specialised LBVS methods in several virtual screening tasks, where molecular shape is an influential factor. 

A possible avenue of future research would be to investigate more sophisticated vectorisations of the Morse transform.
Although we applied the Morse transform to shape-based virtual screening here, our pipeline provides a blueprint for applying the Morse transform to further discrete shape analysis tasks in future work.

\section*{Acknowledgements}

A.M.T. thanks Paul Finn for helpful discussions. A.M.T. thanks the EPSRC Centre for Doctoral Training in \textit{Sustainable Approaches to Biomedical Science: Responsible and Reproducible Research} (grant number EP/S024093/1) for his studentship. V.N.'s work is partially supported by the EPSRC under Grant EP/R018472/1 and in part by US AFOSR under Grant FA9550-22-1-0462. The authors would like to acknowledge the use of the University of Oxford Advanced Research Computing (ARC) facility (\url{http://dx.doi.org/10.5281/zenodo.22558}).

\section*{Data and code availability}

The DUD-E dataset may be downloaded from \url{https://dude.docking.org/} and the MUV dataset is available from \url{https://www.tu-braunschweig.de/en/pharmchem/forschung/baumann/translate-to-english-muv}. 
The underlying code is available at
\url{https://github.com/alex-tanaka/morse-transform-for-drug-discovery}.

\renewcommand{\bibfont}{\small}

\clearpage
\counterwithin{figure}{section}
\counterwithin{table}{section}
\renewcommand\thefigure{\thesection\arabic{figure}}
\renewcommand\thetable{\thesection\arabic{table}}

\appendix

\section{Piecewise-linear Morse theory}
\label{appendix:plmorse}

\begin{figure}[!ht]
\centering
\includegraphics[width=0.8\textwidth]{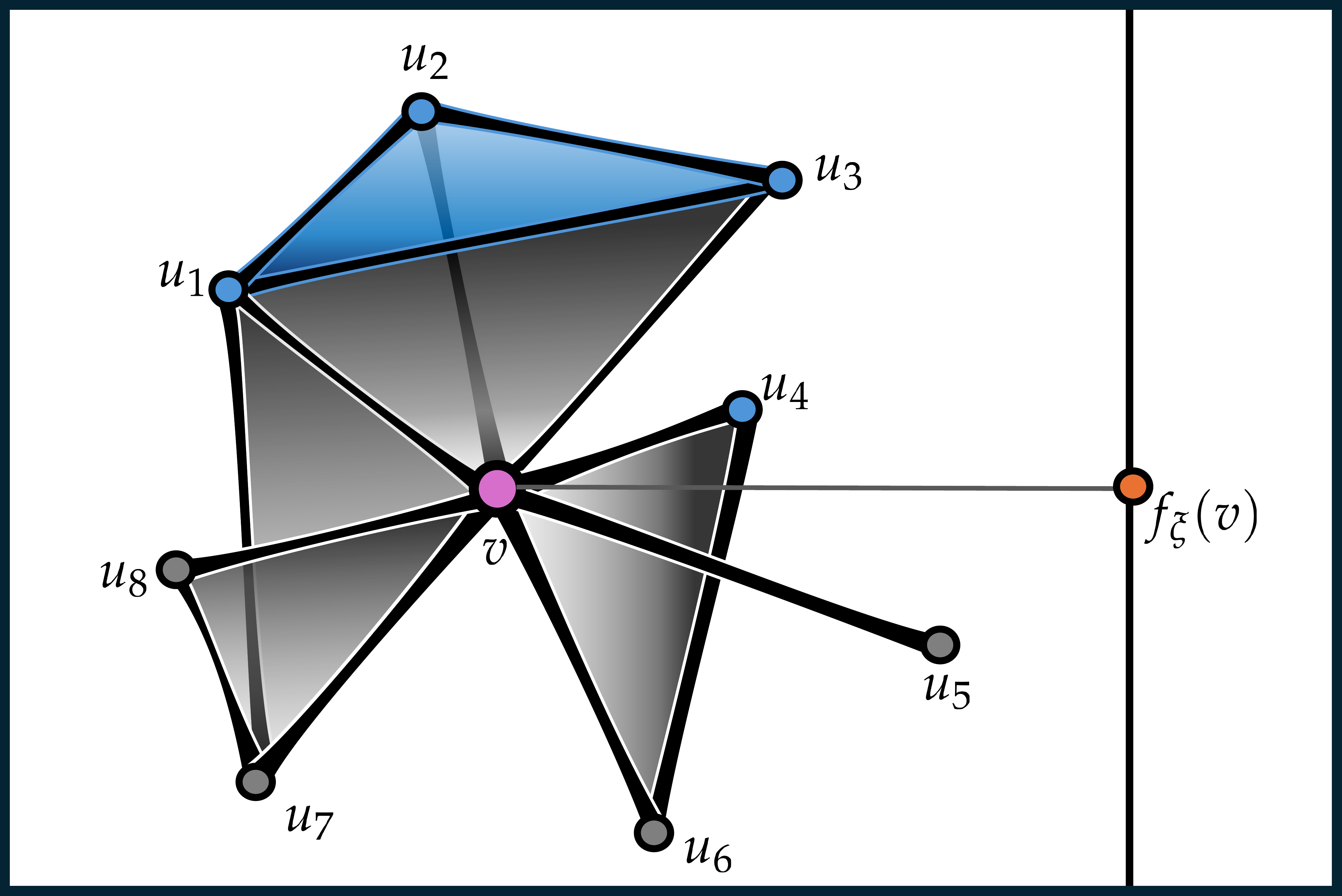}
\caption{The neighbourhood of the vertex $v$ in the illustrated simplicial complex has vertices $\{u_1,\ldots,u_8\}$; its higher-dimensional simplices are $(u_1\,u_2\,u_3\,v)$, $(u_1\,u_7\,v)$, $(u_4\,u_6\,v)$, $(u_5\,v)$, $(u_7\,u_8\,v)$ plus all their faces. The upper link of $v$ with respect to the vertical height-function $f_\xi$ is the blue region containing the 2-simplex $(u_1\,u_2\,u_3)$ plus all its faces along with the isolated vertex $u_4$. The reduced Betti number of the upper link is non-zero in dimension 0; therefore, $v$ is a critical vertex for $f_\xi$. If we consider the same figure rotated clockwise by $90$ degrees so that only $\{u_1, u_2, u_7,u_8\}$ were above $v$, then the upper link would be acyclic and have trivial Betti numbers and $v$ would be non-critical for the corresponding height function.}\label{fig:uplink} 
\end{figure}

Figure~\ref{fig:uplink} is an illustration of how to determine whether or not a vertex is critical based upon its upper link.

\section{Example Morse transform computation}
\label{appendix:mt_example}

\begin{figure}[ht!]
\centering
\includegraphics[width=1.0\linewidth]{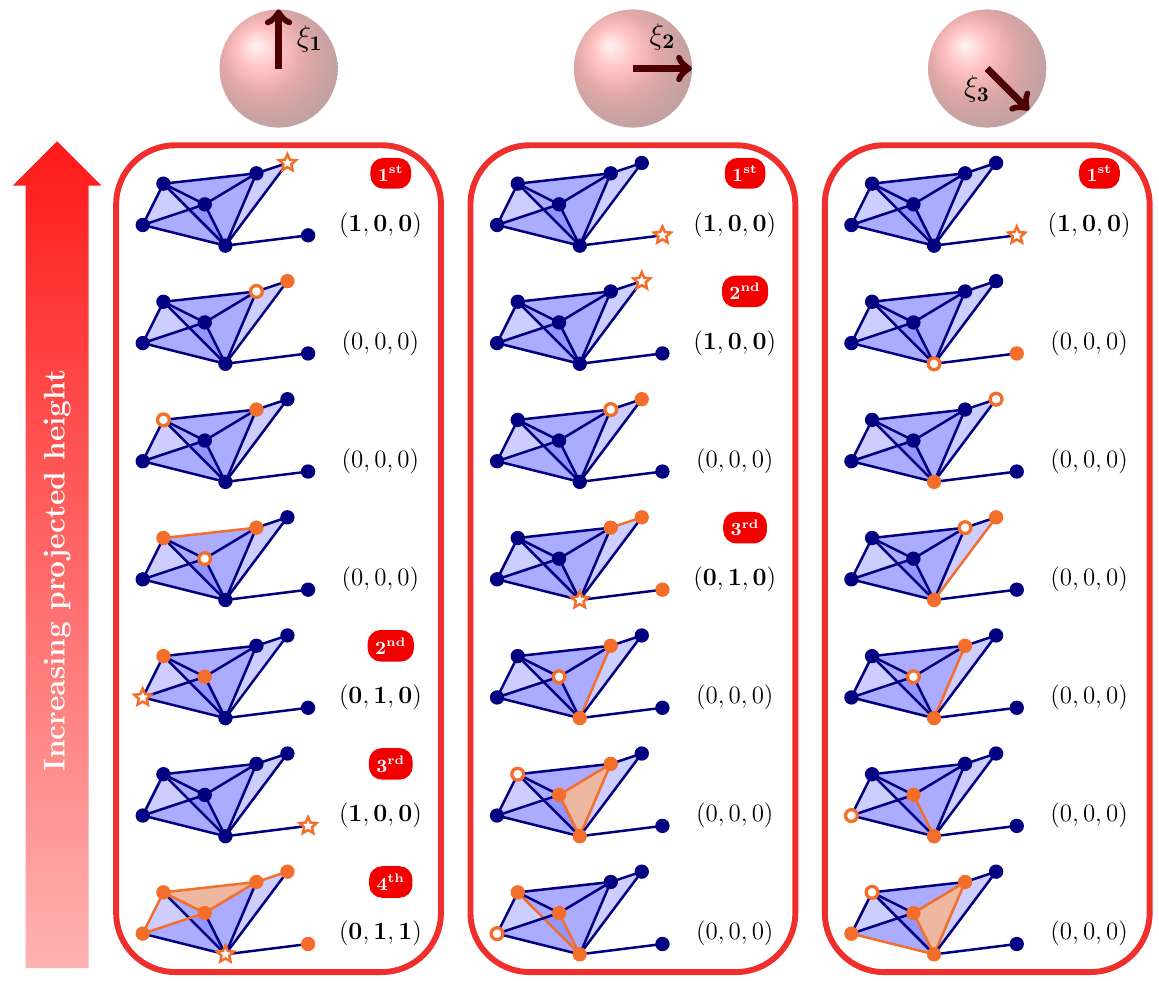}
\caption{Evaluating the critical vertices of a simplicial $3$-complex along three height functions specified by the directions $\xi_1$, $\xi_2$ and $\xi_3$.
For each vertex, we: highlight its upper link; indicate whether it is a critical vertex (a star) or a regular vertex (a circle); write its Morse index to the right; and if its critical, then we label its position in a list of critical vertices ordered by descending height.
(It may appear that some edges of the simplicial $3$-complex intersect at points that are not vertices but this is just an artefact of viewing a 2D projection of the 3D complex.)
}\label{fig:mt_example} 
\end{figure}

Consider a simplicial complex $X$ embedded in $\mathbb{R}^3$ that has $7$ vertices, $12$ edges, $7$ faces and $1$ tetrahedron.
It is akin to a spoon with an antenna.
Suppose that we calculate the Morse transform of $X$ at various depths along three directions $\xi_1, \xi_2, \xi_3 \in \mathbb{S}^2$. 
For ease of illustration, all three directions lie in the 2D plane of the paper.
Figure~\ref{fig:mt_example} displays the process of finding the critical vertices of $X$ along each direction.
Immediately, we can see that there are different numbers of critical vertices along each direction: four along $\xi_1$, three along $\xi_2$ and one along $\xi_3$.
Consequently, the Morse transform at all depths $d \geq 4$ yields exactly the same output of matrices of varying number of rows, corresponding to the number of critical vertices, for each direction.
Note that the fourth critical vertex along $\xi_1$, with a mixed multiplicity Morse index $(0, 1, 1)$, is not Euler-critical so would be undetectable by the Morse Euler-critical transform or the Euler characteristic transform.
At depths $d < 4$, the Morse transform progressively loses information about all the $(d+1)$-th and above critical vertices until at $d=1$ we only retain the highest critical vertex for each direction.
(For a full Morse transform computation we would also need to record the projected heights of each critical vertex.)

\section{The Morse Euler-critical transform}
\label{appendix:met}

Consider a finite simplicial complex $K \subset \mathbb{R}^n$. 
Given a direction vector $\xi$ lying on the unit sphere $\mathbb{S}^{n-1} \subset \mathbb{R}^n$, let $f_\xi:K_0 \to \mathbb{R}$ be the inner-product map $v \mapsto \langle v,\xi \rangle$. 
A vertex $v$ is \textit{Euler-critical} if the Euler characteristic $\chi(\mathrm{Lk}^+(f_\xi;v))$ of the upper link of the vertex is not equal to one.
Let $N_\xi$ be the number of Euler-critical points of $f_\xi$; therefore, we may order the Euler-critical values of $f_\xi$ as $c^\xi_1 > \cdots > c^\xi_{N_\xi}$ and write $v^\xi_i$ for the unique Euler-critical point with value $c^\xi_i \coloneq f_\xi(v^\xi_i)$. 
For brevity, let the Euler characteristic of the upper link $\mathrm{Lk}^+(f_\xi;v^\xi_i)$ be denoted by $\chi(\xi)_i$. 

We define a variant of the Morse transform (MT), the \textit{Morse Euler-critical transform (MET)} of $K$ (of depth $d \in \mathbb{N}$) along $\xi$ to be a function
\begin{align*}
\mathscr{M}^\mathscr{E}_{K,d}: \mathbb{S}^{n-1} &\to \coprod_{k=0}^{d}\mathrm{Mat}_\mathbb{R}(k,2)
\end{align*}
that sends a unit vector $\xi \in \mathbb{S}^{n-1}$ to a certain $m \times 2$ matrix of real numbers, where $m \coloneq \mathrm{min}\{d, N_\xi\}$. 
The $i$-th row of this matrix catalogues the relevant Euler data of $f_\xi$ at the $i$-th Euler-critical point $v_i^\xi$; explicitly, it is given by
\begin{align*}
\mathscr{M}^\mathscr{E}_{K,d}(\xi)_{i} \coloneq \left[~c^\xi_i \quad \chi(\xi)_{i}~\right].
\end{align*}
That is to say, the first column of $\mathscr{M}^\mathscr{E}_{K,d}(\xi)$ contains the top $m$ Euler-critical values of $f_\xi$ in descending order, whilst the remaining column, which is integer-valued, records the Euler characteristic of the corresponding upper links. 
We can also define the \textit{reduced Morse Euler-critical transform (rMET)}
$\mathscr{M}^{\widetilde{\mathscr{E}}}_{K,d}$ just like the MET except all instances of the Euler characteristic are replaced with the reduced Euler characteristic $\widetilde{\chi} \coloneq \chi - 1$ and a vertex is Euler-critical if its upper link has $\widetilde{\chi} \neq 0$.

The MT contains strictly more information than the MET as the Euler characteristic of a simplicial complex $K$ may be calculated from the reduced Betti numbers of the complex via the Euler-Poincar\'e formula
\begin{equation*}
\chi(K) = 1 + \widetilde{\chi}(K) = 1 + \sum_{i=-1}^{d} (-1)^i\, \widetilde{\beta}_i(K) \:.
\end{equation*}
As a consequence, the MT can detect homologically critical points that are otherwise invisible to the MET, as points can be PL Morse-critical but fail to be Euler-critical due to cancellations from reduced Betti numbers of opposite parity.
For example, a point with PL Morse index $\mu = (0, 5, 5)$ is a critical point, but it is not Euler-critical. 

The Morse statistics vectorisation of the MET produces a vector of dimension $27$, which does not depend on the ambient dimension of the simplicial complex.

\section{Comparison to other topological transforms}
\label{appendix:transform_comparisons}

The Morse transform (MT) is a member of a family of topological transforms that produce summaries of shapes by examining them from multiple directions.
The archetypal topological transforms are the Euler characteristic transform (ECT) and the persistent homology transform (PHT) from \citet{pht}. 
The PHT of a finite simplicial complex $K \subset \mathbb{R}^n$ is a function $\mathscr{P}_{K}: \mathbb{S}^{n-1} \to \mathcal{D}^n$ that sends a unit vector of $\xi \in \mathbb{S}^{n-1}$ to a collection of persistence diagrams (PDs). More explicitly,
\begin{equation}
    \mathscr{P}_{K}(\xi) \coloneq \big(\mathrm{PD}_0(K),\, \mathrm{PD}_1(K),\, \dots \mathrm{PD}_{n-1}(K)\big)\:,\nonumber
\end{equation}
where $\mathrm{PD}_i(K)$ is the persistence diagram of $K$ in homological degree $i$ with respect to a height function $f_\xi: K_0 \to \mathbb{R}$ given by $v \mapsto \langle v,\xi \rangle$. Persistent homology computes the lifetime of homology classes (e.g. connected components, tunnels and voids) in families of nested spaces (e.g. a sequence of growing superlevel sets) and persistence diagrams are a two-dimensional summary of persistent homology (see \citet{roadmap} for a more technical explanation of persistent homology). Variants include the extended persistent homology transform \citep{xpht} and the distance-from-flat persistent homology transform \citep{dff_pht}.

Similarly, the  ECT of a simplicial complex $K \subset \mathbb{R}^n$ is a function $\mathscr{E}_{K}: \mathbb{S}^{n-1} \to \mathbb{Z}^\mathbb{R}$ that sends a unit vector of $\xi \in \mathbb{S}^{n-1}$ to a function from $\mathbb{R}$ to $\mathbb{Z}$. More explicitly,
\begin{equation}
    \mathscr{E}_{K}(\xi) \coloneq \mathrm{ECC}_{K,\xi}: \mathbb{R}\to\mathbb{Z} \:,\nonumber
\end{equation}
where $\mathrm{ECC}_{K,\xi} : t \mapsto \chi\big(K_{\leq t}(\xi)\big)$ is the Euler characteristic curve that returns the Euler characteristic $\chi$ of each sublevel set complex $K_{\leq t}(\xi) \coloneq \{\sigma \in K,\, t \in \mathbb{R} \mid f_\xi(v) \leq t \text{ for all } v \in \sigma_0\}$ with respect to the height function $f_\xi$. Some extensions of the ECT are the smooth Euler characteristic transform \citeprefix{SECT;}{sect}, the weighted Euler characteristic transform \citeprefix{WECT;}{wect} and the hybrid transform \citep{hybrid_transform}.

The MT employs a finer topological invariant than the ECT since along one direction the reduced Betti numbers of upperlinks are able to detect homological changes in growing superlevel sets that are otherwise invisible to the Euler curve. 
However, the MT is topologically coarser than the PHT as persistent homology pairs critical points (birth/death points) to compute the lifetime of topological features, unlike the MT.

\subsection{The MT determines the ECT}

We define the upper star $\mathrm{St}^+(f; v)$ of a vertex $v$ in a finite simplicial complex $K$ with respect to a function $f \colon K_0 \to \mathbb{R}$ to be the set of all simplices $\sigma \in K$ that have $v$ as a face and whose vertices $\sigma_0$ have images under $f$ greater than or equal to $f(v)$. 
Note that the upper star is not generally a simplicial complex.
The upper closed star $\overline{\mathrm{St}}^+(f; v)$ of a vertex $v$ in $K$ with respect to $f$ is the closure of the upper star $\mathrm{Cl}\big(\mathrm{St}^+(f; v)\big)$ or equivalently the cone $v \star \mathrm{Lk}^+(f; v)$ of the upper link at $v$. 
Since any such cone is contractible (via the straight-line homotopy to $v$) and a simplicial complex, it follows that the Euler characteristic of every closed star is $1$.

\begin{proposition}
\label{prop:injectivity}
The Morse transform and the Morse Euler-critical transform at full depth 
(equal to the number of vertices of the analysed simplicial complex)
determine the Euler characteristic transform.
\end{proposition}
\begin{proof}
Let $K \subset \mathbb{R}^n$ be a finite simplicial complex. 
Given a direction vector $\xi$ lying on the unit sphere $\mathbb{S}^{n-1} \subset \mathbb{R}^n$, let $f_\xi:K_0 \to \mathbb{R}$ be the inner-product map $v \mapsto \langle v,\xi \rangle$. 
The Euler characteristic transform for one direction is the Euler characteristic curve and may be written in terms of upper links as
\begin{align*}
\mathrm{ECC}_{K,\xi}(t) &\coloneq \chi\big(K^{\geq t}(\xi)\big) && \text{by definition}\\
    &= \sum_{\{v \in K_0 \mid  f_\xi(v)  \geq t\}} \!\!\!\!\!\!\!\! \chi\big( \mathrm{St}^+(f_\xi; v) \big) && \text{by additivity of $\chi$} \\
    &= \sum_{v \in K_0} \chi\big( \mathrm{St}^+(f_\xi; v) \big) \,\mathbb{1}_{ ( -\infty, \langle v, \xi \rangle ] }(t)\\
    &= \sum_{v \in K_0} \Big(1 - \chi\big(\mathrm{Lk}^+(f_\xi; v) \big) \Big) \,\mathbb{1}_{ ( -\infty, \langle v, \xi \rangle ] }(t) \: ,
\end{align*}
where in the last step we have used $1 = \chi(\overline{\mathrm{St}}^+) = \chi(\mathrm{St}^+) + \chi(\mathrm{Lk}^+)$ by combining the contractibility of closed stars with the additivity of Euler characteristics.
Let $\mathrm{Crit}(K; \xi) \subseteq K_0$ be the set of critical points of $K$ for $f_\xi$ and $\mathrm{Crit}^{\mathscr{E}}(K; \xi) \subseteq \mathrm{Crit}(K; \xi)$ be the corresponding set of Euler-critical points. 
Recall that $\widetilde{\beta}(\xi)_j$ is zero whenever $j \geq n-1$, since generically the upper link embeds into $\mathbb{D}^{n-1}$ and all simplicial complexes that embed into $\mathbb{R}^{n-1}$ have vanishing $k$-dimensional singular homology groups for $k \geq n-1$.
We have by the definition of critical points, the Euler-Poincar\'e formula and the definition of the Morse transform that 
\begin{align*}
\mathrm{ECC}_{K,\xi}(t) 
    &= \sum_{v \in \mathrm{Crit}^{\mathscr{E}}(K; \xi)} \!\!\Big(1 - \chi\big(\mathrm{Lk}^+(f_\xi; v) \big) \Big) \,\mathbb{1}_{ ( -\infty, \langle v, \xi \rangle ] }(t)\\
    &= \sum_{v \in \mathrm{Crit}(K; \xi)} \,\sum_{j=-1}^{n-2} (-1)^{j+1}\, \widetilde{\beta}_j\big(\mathrm{Lk}^+(f_\xi; v) \big) \,\mathbb{1}_{ ( -\infty, \langle v, \xi \rangle ] }(t)\\
    &= \sum_{v \in \mathrm{Crit}(K; \xi)} \,\sum_{j=-1}^{n-2} (-1)^{j+1}\, \mathscr{M}_{K,d}(\xi)_{i(v), j+3} \,\mathbb{1}_{ \big( -\infty, \,\mathscr{M}_{K,d}(\xi)_{i(v), 1} \big] }(t) \: ,
\end{align*}
where 
$i(v) \coloneq 1 + |\{w \in \mathrm{Crit}(K; \xi) \mid \langle w,\xi\rangle > \langle v,\xi\rangle \}|$ gives the position of $v$ in a list of critical points when ordered by descending images under $f_\xi$, or equivalently the row of the Morse transform corresponding to $v$. 
Therefore, the Morse transform at full depth determines the Morse Euler-critical transform at full depth, which in turn completely determines the Euler characteristic transform.
\end{proof}
This also provides an alternative way of computing the ECT via the MET, which is similar to  \citet{euler_critical_ect_vadim_2024} except our construction computes the Euler characteristic through the upper link of the vertex rather than its upper star.
We use this method for generating the ECT comparisons.

\subsection{Injectivity}

The most celebrated property of the PHT and ECT is that they are injective on the space of finite simplicial complexes embedded in $\mathbb{R}^3$ \citep{pht} and more generally constructible sets in $\mathbb{R}^n$ \citep{pht_how_many_directions, pht_injectivity_ghrist}. 
Strikingly, any shape from a reasonably behaved uncountable set of shapes can be uniquely determined from only a finite number of directions using the PHT or ECT \citep{pht_how_many_directions}. 
By Proposition~\ref{prop:injectivity}, the MT and MET at full depth completely determine the ECT.
Hence, the MT and MET inherit the injectivity of the ECT. 
Whether the MT requires fewer or more directions on average than the ECT or PHT to perfectly reconstruct a simplicial complex (theoretically or in practice) is an open question.

\subsection{Discretisation along one direction}

The MT has a different approach to discretisation along one direction than the ECT. Typically, discretising the ECT along each
direction involves selecting a finite set of height thresholds at which to evaluate the ECC. These are often chosen to be uniformly distributed over a certain range of heights that cover the whole shape. In contrast, the MT requires selecting an integer depth in advance that restricts the number of critical vertices (in descending height) examined in each direction up to that depth. Consequently, the MT prioritises shape information closer to the outer surface of the shape unlike the ECT. This is especially relevant for protein-ligand interactions, which depend on the shape of the surface of the ligand. 
Furthermore, by definition, the MT looks at critical vertices where the homology of the superlevel sets changes, but usually ECT height thresholds are not chosen to coincide with where the Euler characteristic changes (at \textit{Euler-critical} points). 
Only recently has an implementation of the ECT been developed to reduce computational complexity by focusing on Euler-critical points of weighted cubical complexes \citep{euler_critical_ect_vadim_2024}.

\subsection{Vectorisation}

Most machine learning algorithms expect vectors as inputs; therefore, it is of practical importance to have a way of converting the output of topological transforms to vectors.
The PHT may be vectorised by concatenating the individual vectorisations of the component PDs for each direction. There are many different vectorisations of PDs to choose from, such as persistence landscapes, persistence images, Betti curves and persistence statistics, and each has its own strengths \citep{pd_vectorizations}. The ECT may be vectorised by flattening an Euler characteristic profile \citep{euler_profile}, which is a matrix where each column consists of images of the height thresholds under an ECC for a certain direction. 
We provide a simple vectorisation for the MT: Morse statistics, which is compact and whose length is independent of the number of directions examined and their labelling. 
The downside of this vectorisation scheme is that it obfuscates the explicit directional information associated to each critical vertex as it only summarises the distribution of local Morse data.

\clearpage
\section{Rotational invariance}
\label{appendix:rot_inv}

The Morse transform $\mathscr{M}_{K,d}(\xi)$ (of depth $d$) of a simplicial $n$-complex $K_w(P')$ along a direction $\xi \in \mathbb{S}^{n-1}$ is not intrinsically invariant under the action of the special orthogonal group $\mathrm{SO}(n)$ on the underlying point cloud $P' \subset \mathbb{R}^n$ unless $P'$ only consists of one point at the origin. However, the Morse feature vector of $P'$ is invariant under the action of $\mathrm{SO}(n)$ on $P'$ owing to the intermediate step of orienting $P'$, as described below.

Let $C = \mathrm{cov}(P', P')$ be the covariance matrix of $P'$ and its eigendecomposition be $C = E \Lambda E^\mathrm{T}$, where $\Lambda = \mathrm{diag}(\lambda_1, \lambda_2, \dots, \lambda_n)$ is the matrix of eigenvalues (\textit{principal component values}) of $C$ and  $E = \big[e^{(1)},\, e^{(2)}, \,\dots\, ,\, e^{(n)}\big]$ is the matrix of the associated eigenvectors (\textit{principal component axes}). Then the point cloud $P'$ is oriented by first aligning $P'$ with the principal component axes and thus transforming it into a \textit{canonical pose} $P_c = P'E$, which is invariant under the action of $\mathrm{SO}(n)$ on $P'$. Due to the lack of uniqueness of the eigendecomposition, there is more than one canonical pose as there are $n!$ ways of ordering the eigenvectors in $E$ and reflections of the eigenvectors $\pm e^{(i)}$ also satisfy the eigendecomposition leading to $n!\cdot2^n$ canonical poses \citep{pca_ambiguity}. A \textit{primary pose} $P$ is chosen from the set of canonical poses $\{P_c\}$ by 
\begin{enumerate}
    \item ordering the eigenvalues such that $\lambda_1 > \lambda_2 > ... > \lambda_n$, leaving $2^n$ canonical poses (assuming no repeated eigenvalues);
    \item selecting the signs of the eigenvectors such that $\mathrm{skew}[P'e^{(i)}] > 0$ for all $i < n$, leaving two canonical poses (assuming that $P'$ does not have zero skew along any principal component axis); and 
    \item choosing the sign of the last eigenvector $e^{(n)}$ such that $|E| = 1$ (making $E$ a proper rotation), leaving one canonical pose.
\end{enumerate}
In this way, we can orient a point cloud $P'$ into the primary pose $P$ in a manner that is invariant under rotations of $P'$. In turn, the Morse feature vector that is derived from the primary pose is also rotationally invariant. (See Appendix~\ref{appendix:pca_analysis} for a qualitative study on how likely the PCA-based alignment scheme is to succeed for the two datasets used in the paper.)

\section{Dataset PCA analysis}
\label{appendix:pca_analysis}

\begin{figure*}[ht!]
    \centering
    \begin{subfigure}{0.8\linewidth}
    \centering
    \includegraphics[width=\linewidth]{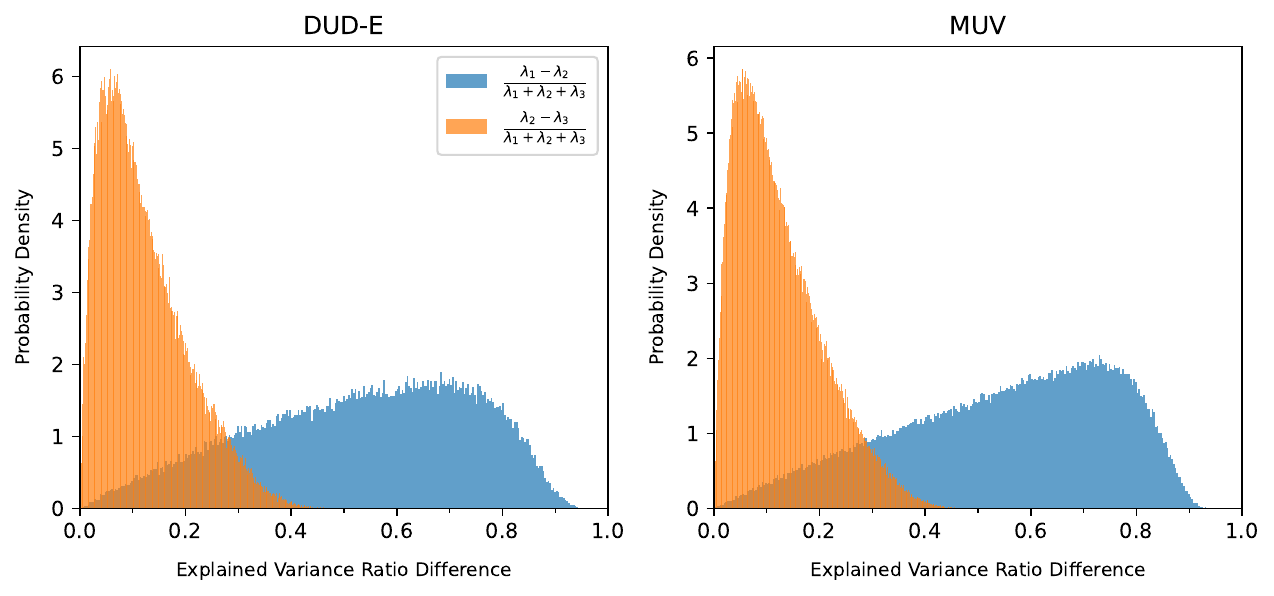}
    \caption{}
    \label{subfig:pca_differences}
    \end{subfigure}\\
    \begin{subfigure}{0.8\linewidth}
    \centering
    \includegraphics[width=\linewidth]{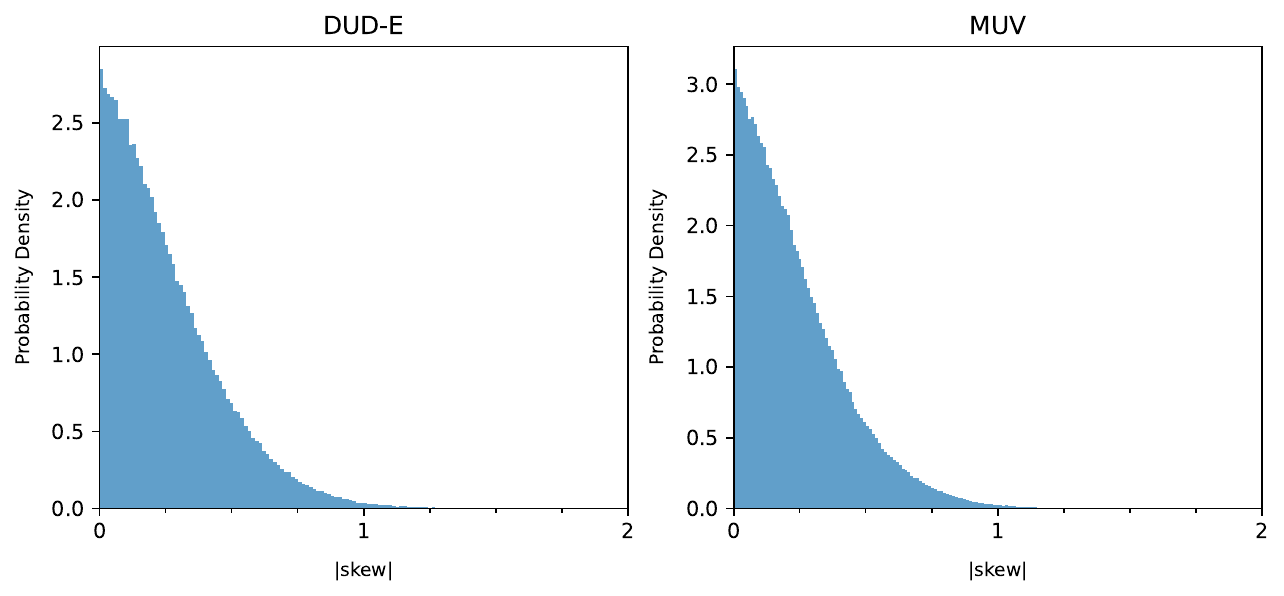}
    \caption{}
    \label{subfig:skews}
    \end{subfigure}\\  
    \caption{Panel~\ref{subfig:pca_differences} depicts histograms of the normalised differences between the $1^{\mathrm{st}}-2^{\mathrm{nd}}$ and the $2^{\mathrm{nd}}-3^{\mathrm{rd}}$ largest principal component values of the molecular point clouds in DUD-E and MUV. Panel~\ref{subfig:skews} depicts histograms of the absolute values of the skews of the molecular point clouds along their principal component axes in DUD-E and MUV.}
  \label{fig:pca_analysis}
\end{figure*}

Our Principal Component Analysis (PCA) based alignment scheme relies on the principal component values of each point cloud being non-degenerate and thus possessing a strict total order. We plot histograms of the differences between the principal component values (normalised by the sum of the principal component values) in Fig.~\ref{subfig:pca_differences} and the magnitude of the skews along the principal component axes of each point cloud in Fig.~\ref{subfig:skews} for the two datasets DUD-E and MUV to qualitatively asses how likely the alignment scheme is to succeed. 
The distribution of the normalised difference between the $1^{\mathrm{st}}-2^{\mathrm{nd}}$ largest principal components peaks at a high difference, tends towards zero as the difference approaches zero and is skewed towards lower values. The distribution for the normalised difference between the $2^{\mathrm{nd}}-3^{\mathrm{rd}}$ largest principal components peaks at a lower difference, also tends towards zero as the difference approaches zero and is skewed towards higher values. None of the normalised differences between principal components are less than $1 \times 10^{-5}$. This implies that the molecules are generally cylindrically shaped but with enough asymmetry to impose a total order on the principal components. 

The distribution of the absolute skews along the principal component axes is peaked around zero skew, skewed towards higher values and is not too sharply concentrated. None of the absolute skews are less than $1 \times 10^{-7}$. This means that it is possible to uniquely orient the point cloud along each principal component axis by examining the skews. However, this skew-based orientation scheme may be sensitive to noise as a number of skews are close to zero and conceivably could be perturbed to being indistinguishable from zero under $32$-bit floating point precision.

\clearpage
\section{Chemical properties}
\label{appendix:virtual}

To a large extent, the biological activity of a ligand depends upon how well it binds to a target protein \citep{dicera2020mechanisms}; this binding affinity is a complicated interplay between the geometric and chemical properties of both molecules. 
We encode the geometric properties of molecules using a novel method based upon PL Morse theory, but the chemical information we incorporate is more conventional. Besides knowledge of the atom locations in 3D and their van der Waals radii, we only make use of the following chemical properties: 
\begin{enumerate}
    \item the MMFF94 modified \textit{partial charge} as implemented in MOE \citep{moe}, which accounts for the asymmetric distribution of electrons in the chemical bonds of a molecule; 
    \item the atomic contribution to Wildman-Crippen \textit{molar refractivity} \citep{wildman_crippen}, which is a measure of the polarizability of the molecule; and
    \item the atomic contribution to the Wildman-Crippen \textit{lipophilicity} \citep{wildman_crippen}, which measures the equilibrium distribution of the molecule between a non-polar and a polar solvent.
\end{enumerate}
In particular, we do not assume any prior knowledge of the geometric or chemical properties of the target protein binding site, unlike in structure-based virtual screening.

\section{Hyperparameter tuning}
\label{appendix:tuning}

We tune the hyperparameters of the LGBM classifier using a random search of $200$ samples of the hyperparameter search space given in Table~\ref{tab:tuning}. For each $5$-fold cross-validation split of the data, the best performing hyperparameters are selected by choosing the model with the lowest mean log loss across an \textit{inner} $4$-fold cross-validation of the training subset of the data. Then the best hyperparameters are used to train the model using the whole training subset of the data, which is then assessed on the held-out test set. This whole process is repeated across each split of the $5$-fold cross-validation. A random seed of $2023$ is used for the $5$-fold cross-validation.

\begin{table}[H]
\centering
\caption{LightGBM hyperparameters and their tuning search space. Given a set of hyperparameters $\mathcal{S}$, then $\text{uniform}(\mathcal{S})$ denotes uniform random sampling of $\mathcal{S}$ and similarly $\text{loguniform}(\mathcal{S})$ denotes uniform logarithmic random sampling of $\mathcal{S}$.}
\label{tab:tuning}
\begin{tabular}{cc}
\toprule
Hyperparameter & Search Space\\
\midrule
bagging\_fraction & $\mathrm{uniform}([0.3, 1.0])$\\
feature\_fraction & $\mathrm{uniform}([0.3, 1.0])$\\
max\_depth & $\mathrm{uniform}([2, 100] \cap \mathbb{N})$\\
min\_data\_in\_leaf &  $\mathrm{loguniform}([20, 2000] \cap \mathbb{N})$\\
min\_sum\_hessian\_in\_leaf & $\mathrm{loguniform}([10^{-5}, 20])$\\
num\_leaves & $\mathrm{loguniform}([2, 4095] \cap \mathbb{N})$\\
l1 & $\mathrm{loguniform}([10^{-5}, 10^{7}])$\\
l2 & $\mathrm{loguniform}([10^{-5}, 10^{7}])$\\
\bottomrule
\bottomrule
\end{tabular}
\end{table}

\clearpage
\section{Evaluation metrics}
\label{appendix:metrics}

The receiver operating characteristic (ROC) curve of a binary classifier is the plot of the true positive rate (TPR) against the false positive rate (FPR) as the threshold of the binary classifier varies. A common evaluation metric is the area under the ROC curve (AUROC) given by
\begin{align}
    \mathrm{AUROC} = \int_{-\infty}^{+\infty}\mathrm{TPR} \,\frac{\mathrm{d (FPR)}}{\mathrm{d}T}\mathrm{d}T\ \in [0,1]\:.
    \label{eq:auroc_continuous}
\end{align}
It measures how likely it is for a randomly selected member of the positive class to be ranked above a randomly selected member of the negative class. A score of 0.5 indicates that the classifier has equivalent performance to a random classifier. 

The Boltzmann-enhanced discrimination of the receiver operating characteristic at $\alpha$ \citeprefix{BEDROC\textsubscript{$\alpha$};}{bedroc} is given by
\begin{align}
    \mathrm{BEDROC\textsubscript{$\alpha$}} = \frac{\mathrm{wAUAC} - \mathrm{wAUAC}_\mathrm{min}}{\mathrm{wAUAC}_\mathrm{max} - \mathrm{wAUAC}_\mathrm{min}}\:,
    \label{eq:bedroc}
\end{align}
where the weighted area under the accumulation curve (wAUAC) is given by
\begin{align}
    \mathrm{wAUAC} = \frac{\int_0^1 F(x)w(x)\mathrm{d}x}{\int_0^1 w(x)\mathrm{d}x} \quad\text{with}\quad w(x) = e^{-\alpha x} 
    \label{eq:wauac}
\end{align}
and $F(x)$ is the empirical cumulative distribution function (CDF) of the positive class. BEDROC\textsubscript{$\alpha$} is similar to the AUROC, except the contribution of the earlier part of the ROC curve is exponentially weighted to have a higher contribution. BEDROC\textsubscript{$\alpha$} is equal to 0.5 if the observed empirical CDF has the shape of the CDF produced by a probability density function proportional to the earlier exponential weight function with parameter $\alpha$.

The enrichment factor at a fraction $\chi \in [0,1]$ (EF\textsubscript{$\chi$}) is given by
\begin{align}
    \mathrm{EF\textsubscript{$\chi$}} = \frac{\int_0^1 F(x)w(x)\mathrm{d}x}{\int_0^1 w(x)\mathrm{d}x} \quad\text{with}\quad w(x) = 
    \begin{cases}
        1 & \text{if } x \leq \chi\\
        0 & \text{if } x > \chi
    \end{cases}
    \quad .
    \label{eq:ef}
\end{align}
It is a popular metric in virtual screening and measures how much more likely it is to find a member of the positive class in the first portion of a ranked sample than in the whole sample.

The relative enrichment factor at a fraction $\chi \in [0,1]$  (REF\textsubscript{$\chi$}) is given by
\begin{align}
    \mathrm{REF\textsubscript{$\chi$}} = \frac{\mathrm{EF\textsubscript{$\chi$}}}{\mathrm{EF\textsubscript{$\chi$, max}}} \: ,
    \label{eq:ref}
\end{align}
where EF\textsubscript{$\chi$, max} is the maximum possible enrichment factor in the first $\chi$ of a ranked sample, which depends on the ratio of the positive to negative class.

\clearpage
\section{Visualisation}
\label{appendix:visualisation}

\begin{figure}[!ht]
    \centering
    \begin{subfigure}{0.48\linewidth}
    \centering
    \includegraphics[width=\linewidth]{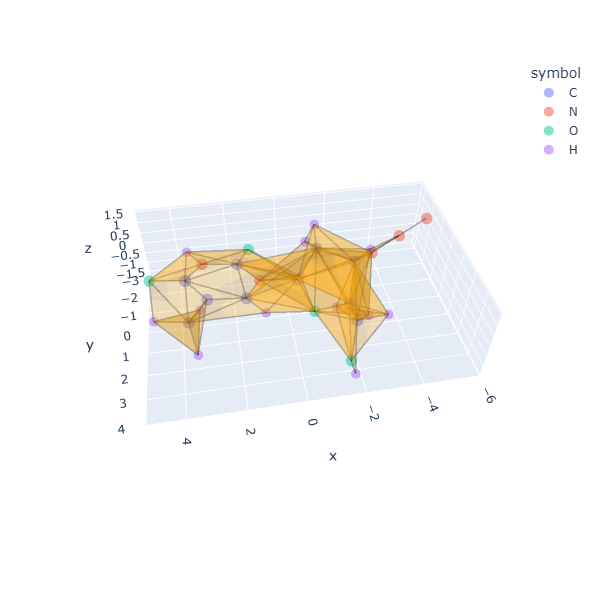}
    \caption{Chemical symbol labelling.}
    \label{subfig:chem_label}
    \end{subfigure}
    \begin{subfigure}{0.48\linewidth}
    \centering
    \includegraphics[width=\linewidth]{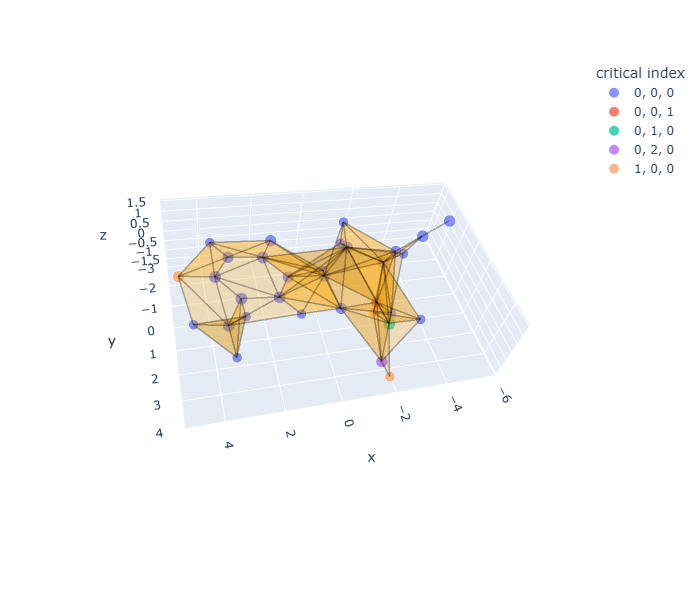}
    \caption{PL Morse index labelling.}
    \label{subfig:index_label}
    \end{subfigure}\\  
    \begin{subfigure}{0.48\linewidth}
    \centering
    \includegraphics[width=\linewidth]{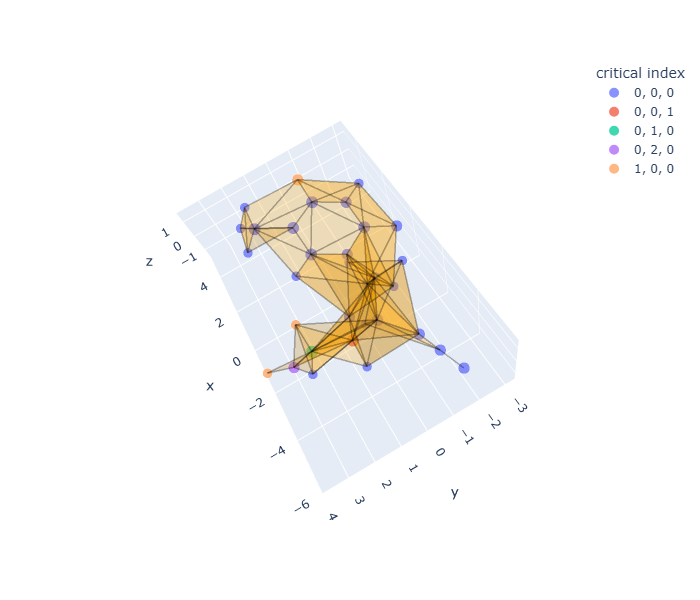}
    \caption{Second rotated view.}
    \label{subfig:index_label_view_2}
    \end{subfigure}
    \begin{subfigure}{0.48\linewidth}
    \centering
    \includegraphics[width=\linewidth]{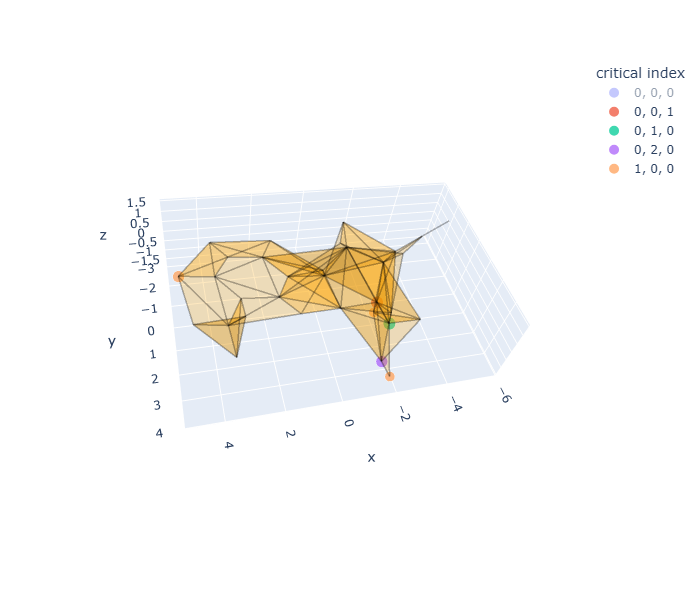}
    \caption{Regular vertices hidden.}
    \label{subfig:index_label_reg_hidden}
    \end{subfigure}\\  
    \caption{Various visualisations of a 3D molecular simplicial complex. The PL Morse height-function is along the $x$-axis.}
  \label{fig:complex_visualisation}
\end{figure}

Our code can visualise 3D molecular simplicial complexes generated from \linebreak pruned Delaunay triangulations.
We have two ways of visualising a molecular simplicial complex: (i) with the vertices labelled by their corresponding chemical elements and (ii) with the vertices labelled by their PL Morse index for a specific height function.
For each type of visualisation, it is possible to zoom, rotate and move the simplicial complex. 
Vertices of specific types may be hidden by selecting the corresponding item in the legend.
Figure~\ref{fig:complex_visualisation} displays a few visualisations of a ligand of the DUD-E target AMPC.

\section{ECT vectorisations}
\label{appendix:ect_vec}

We vectorise the ECT using a procedure that is similar to the most common way of vectorising the ECT. 
For each of the $N_d$ directions in the ECT, we calculate a Euler characteristic curve using $N_t$ thresholds evenly distributed between $[r, -r]$, where $r$ is the maximum norm of any point in the particular simplicial complex.  
Then we concatenate all of the Euler characteristic curves along with one copy of the thresholds together to give a vector of dimension $(N_d + 1) N_t$.
In the main text we refer to this vectorisation as \textit{ECT-F}.
Note that the order of the concatenated Euler characteristic curves in the concatenation is arbitrary as it depends upon the arbitrary ordering of the analysed directions.

We also implement another way of vectorising the ECT, \textit{ECT statistics}, which involves taking the $p$-th percentiles of the Euler characteristic for all $p$ in $\{0,\allowbreak 10, \allowbreak 25, \allowbreak 40, \allowbreak 50, \allowbreak 60, 75, 90, 100\}$ (the same ones in our Morse statistics vectorisation). 
We then concatenate all the percentiles along with $r$ to encode the scale of the simplicial complex.
This gives a vector of length $10$.
In the main text we refer to this vectorisation by \textit{ECT-P}.
This vectorisation is invariant to the arbitrary labelling of directions.

\section{Comparison with superpositional methods}
\label{appendix:similarity}

Superpositional methods rank molecules by their similarity to a reference or template molecule using a hand-crafted similarity scoring function. In Table~\ref{tab:dude_auroc_sim} we record the mean AUROC of similarity ligand-based virtual screening methods for DUD-E. We reproduce the results for eSim from \citep{esim}; USR, USRCAT, ROCS (shape), ROCS (colour) and VAMS from \citep{vams}; CDK-D Moments, Shape-IT and Spectral Geometry Covariance 100 evaluations from \citep{spectral_geometry_feature}; Interlig from \citep{interlig}; and Optipharm-Robust and WEGA from \citep{optipharm}.

In Table~\ref{tab:muv_auroc_sim} we record the mean AUROC of similarity ligand-based virtual \linebreak screening methods for MUV. 
We reproduce the results for VSFlow (shape COMBO) from \citep{vsflow}; Interlig from \citep{interlig}; and BRUTUS and ROCS (colour) from \citep{comp_muv_tiikkainen}.

Generally, similarity methods perform worse than machine learning models trained on non-superpositional descriptors, which can be seen by comparing Tables~\ref{tab:dude_auroc_sim} and \ref{tab:muv_auroc_sim} with Table~\ref{tab:dude_muv_auroc} in the main text.

\begin{table}[ht!]
\centering
\caption{The mean AUROC per DUD-E Diverse (D8) subset target and the overall mean AUROC of the full set of 102 DUD-E (D102) targets (if available) for various similarity methods. For each target the method with the highest AUROC is displayed in bold and the highest shape-based method is underlined.
}
\label{tab:dude_auroc_sim}
\footnotesize
\begin{tabular}{lcccccccccccc}
\toprule 
& \multicolumn{8}{c}{Shape} & \multicolumn{4}{c}{Shape \& Chemistry}\\
\cmidrule(lr){2-9}\cmidrule(lr){10-13}
Target & \rothead{USR} & \rothead{CDK-D} & \rothead{Optipharm-\\Robust} & \rothead{ROCS (shape)} & \rothead{SGC} & \rothead{Shape-IT} & \rothead{VAMS} & \rothead{WEGA} & \rothead{eSim} & \rothead{Interlig} & \rothead{ROCS (colour)} & \rothead{USRCAT}\\
\midrule
AKT1 & $0.36$ & $0.58$ & $0.26$ & $0.28$ & $\underline{0.62}$ & $0.64$ & $0.41$ & $0.26$ & $0.58$ & $\bf 0.67$ & $0.37$ & $0.40$\\
AMPC & $0.53$ & $0.63$ & $0.63$ & $0.58$ & $0.58$ & $0.59$ & $\underline{0.64}$ & $\underline{0.64}$ & $0.62$ & $\bf 0.76$ & $\bf 0.76$ & $0.73$\\
CP3A4 & $0.53$ & $0.52$ & $0.53$ & $\underline{0.55}$ & $0.51$ & $0.54$ & $0.54$ & $0.53$ & $0.58$ & $\bf 0.70$ & $0.51$ & $0.51$\\
CXCR4 & $0.62$ & $0.67$ & $0.71$ & $\underline{0.78}$ & $0.66$ & $0.65$ & $0.72$ & $0.73$ & $0.79$ & $\bf 0.92$ & $0.78$ & $0.65$\\
GCR & $0.50$ & $0.63$ & $0.52$ & $0.49$ & $0.66$ & $\bf \underline{0.77}$ & $0.49$ & $0.50$ & $0.64$ & $0.76$ & $0.59$ & $0.63$\\
HIVPR & $0.74$ & $0.62$ & $0.70$ & $0.72$ & $0.63$ & $0.62$ & $\underline{0.78}$ & $0.71$ & $\bf 0.84$ & $\bf 0.84$ & $0.69$ & $0.73$\\
HIVRT & $0.62$ & $0.50$ & $0.52$ & $0.63$ & $0.50$ & $0.46$ & $\underline{0.69}$ & $0.52$ & $\bf 0.71$ & $0.67$ & $0.61$ & $0.67$\\
KIF11 & $0.61$ & $0.67$ & $\bf \underline{0.83}$ & $0.76$ & $0.65$ & $0.77$ & $0.68$ & $\bf \underline{0.83}$ & $0.73$ & $0.80$ & $0.76$ & $0.69$\\
\midrule
mean (D8) & $0.56$ & $0.60$ & $0.59$ & $0.60$ & $0.60$ & $\underline{0.63}$ & $0.62$ & $0.59$ & $0.69$ & $\bf 0.77$ & $0.63$ & $0.62$\\
mean (D102) & $0.52$ & $0.58$ & $0.56$ & $0.60$ & 
$-$ & $\underline{0.61}$ & $0.56$ & $0.56$ & $0.76$ & $\bf 0.78$ & $0.66$ & $0.55$\\
\bottomrule
\bottomrule
\end{tabular}
\end{table}
\begin{table}[ht!]
\centering
\caption{The mean AUROC per MUV target of various similarity methods. An asterisk * indicates that the AUROC values were estimated from a figure. For each target the method with the highest AUROC is displayed in bold.
}
\label{tab:muv_auroc_sim}
\footnotesize
\begin{tabular}{lcccc}
\toprule
& \multicolumn{1}{c}{Shape} & \multicolumn{3}{c}{Shape \& Chemistry}\\
\cmidrule(lr){2-2} \cmidrule(lr){3-5}
Target & \multicolumn{1}{p{1.2cm}}{VSFlow (shape COMBO)} & Interlig & BRUTUS & \multicolumn{1}{p{0.975cm}}{ROCS (colour)}\\
\midrule
466 & *$\bf 0.62$ & $0.57$ & $0.51$ & $0.56$\\
548 & *$\bf 0.70$ & $0.65$ & $0.61$ & $0.69$\\
600 & *$\bf 0.59$ & $\bf 0.59$ & $0.50$ & $0.58$\\
644 & *$\bf 0.62$ & $0.58$ & $0.57$ & $0.59$\\
652 & *$0.57$ & $\bf 0.61$ & $0.42$ & $0.57$\\
689 & *$0.53$ & $\bf 0.58$ & $0.46$ & $0.49$\\
692 & *$0.63$ & $0.56$ & $0.63$ & $\bf 0.65$\\
712 & *$0.56$ & $\bf 0.64$ & $0.45$ & $0.58$\\
713 & *$\bf 0.51$ & $\bf 0.51$ & $0.41$ & $0.48$\\
733 & *$\bf 0.54$ & $0.50$ & $0.48$ & $0.53$\\
737 & *$\bf 0.65$ & $0.59$ & $0.46$ & $0.53$\\
810 & *$0.54$ & $\bf 0.65$ & $0.42$ & $0.46$\\
832 & *$\bf 0.70$ & $0.65$ & $0.55$ & $0.62$\\
846 & *$0.70$ & $0.69$ & $0.65$ & $\bf 0.71$\\
852 & *$\bf 0.78$ & $0.67$ & $0.63$ & $0.73$\\
858 & *$0.54$ & $\bf 0.60$ & $0.46$ & $0.52$\\
859 & *$0.49$ & $\bf 0.59$ & $0.51$ & $0.51$\\
\midrule
mean & *$0.60$ & $\bf 0.64$ & $0.51$ & $0.58$\\
\bottomrule
\bottomrule
\end{tabular}
\end{table}

\clearpage
\section{Further ablation studies}
\label{appendix:ablation}

\begin{figure}[H]
\centering
\includegraphics[width=0.45\textwidth]{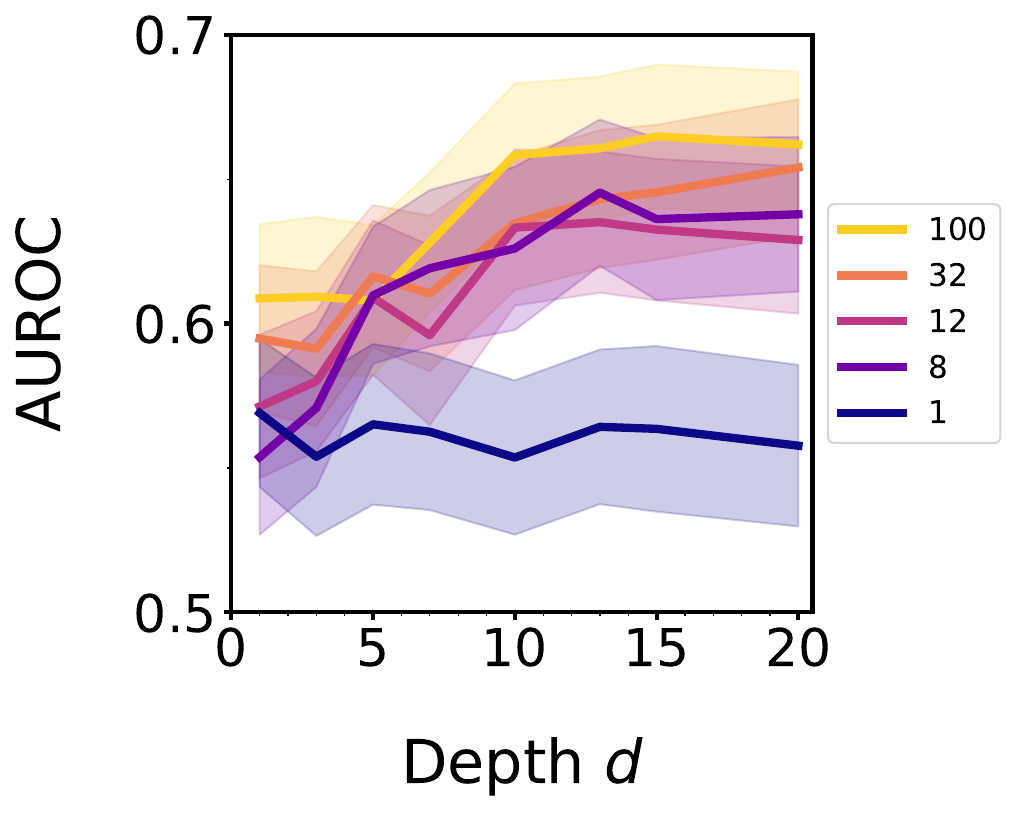}
\caption{The mean AUROC score against depth of our LGBM classifier trained on Morse feature vectors computed using 100 directions (100), 32 pentakis dodecahedral directions (32), 12 icosahedral directions (12), 8 cubic directions (8) and 1 direction (1) for the MUV dataset. The error bars correspond to $95\,\%$ confidence intervals. 
}
\label{fig:muv_auroc_directions_plot}
\end{figure}

\begin{figure}[!ht]
\centering
\includegraphics[width=0.45\textwidth]{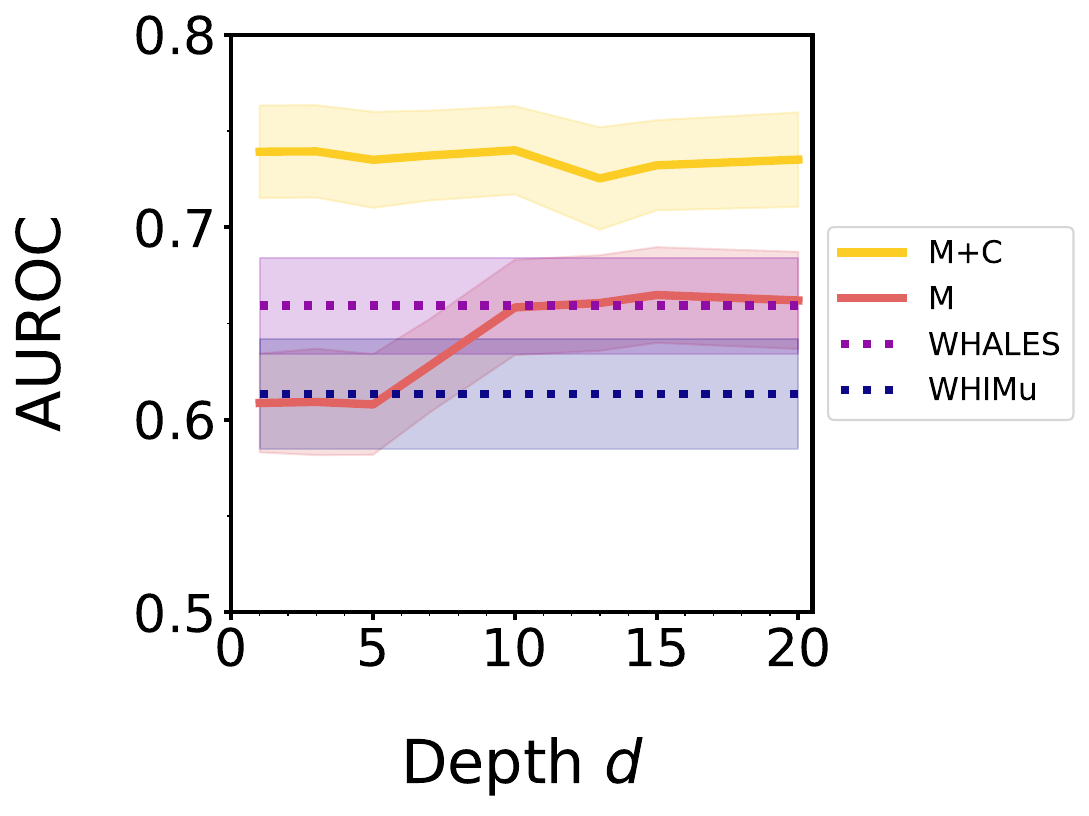}
\caption{The mean AUROC score against depth of the LGBM classifier trained on chemistry-supplemented Morse feature vectors computed using 100 directions (M+C) and Morse feature vectors computed using 100 directions (M) for the MUV dataset. For comparison, the best-performing external shape- and chemistry-based descriptor WHALES (WHALES) and best performing shape-based descriptor WHIMu (WHIMu) are plotted with dotted lines. The error bars correspond to $95\,\%$ confidence intervals. 
}
\label{fig:muv_external_comparison_line_plot}
\end{figure}

In Fig.~\ref{fig:muv_auroc_directions_plot} we show the performance of various Morse features using different numbers of directions for MUV. The performance reaches diminishing returns around 32 directions (though there is a larger difference between 32 and 100 directions than in D8) and depth 15 though the curves are less smooth with more uncertainty. In Fig.~\ref{fig:muv_external_comparison_line_plot} we compare chemistry-supplemented and plain Morse features against the best performing external descriptors in their category for MUV. Similar to D8, chemistry-supplemented Morse features are much less affected by depth.


\begin{figure}[!ht]
\centering
\includegraphics[width=\textwidth]{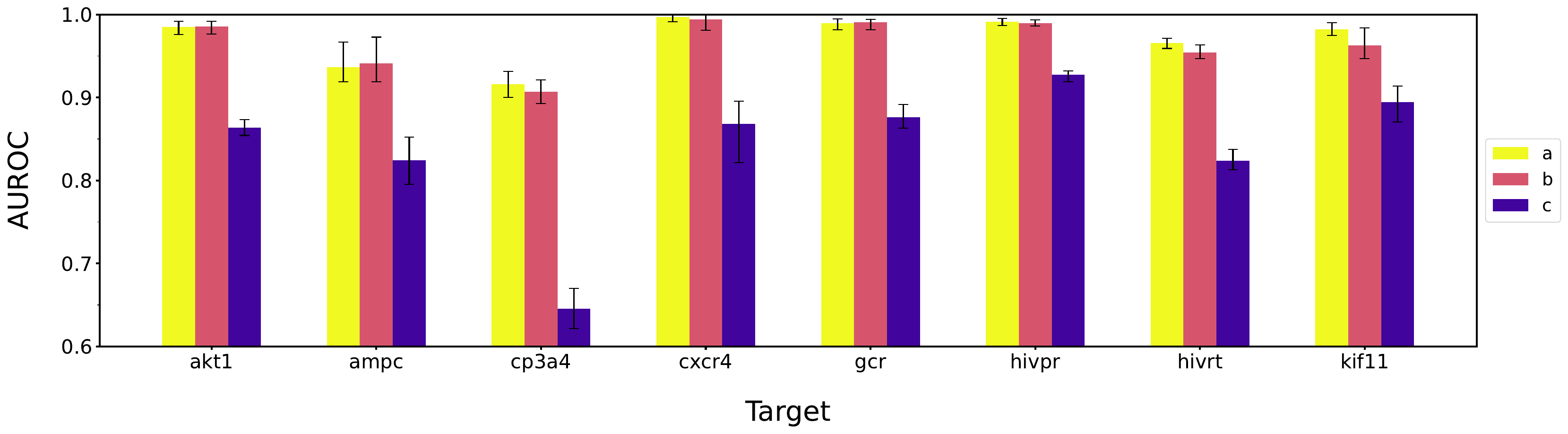}
\caption{The mean AUROC score per DUD-E Diverse target of our classifier trained on Morse features at depth 3 and 32 directions supplemented with the chemical property percentiles (a), chemical property percentiles alone (b), and Morse features at depth 20 and 32 directions (c). The error bars correspond to $95\,\%$ confidence intervals. 
}
\label{fig:dude_diverse_auroc_bar}
\end{figure}

\begin{figure}[!ht]
\centering
\includegraphics[width=\textwidth]{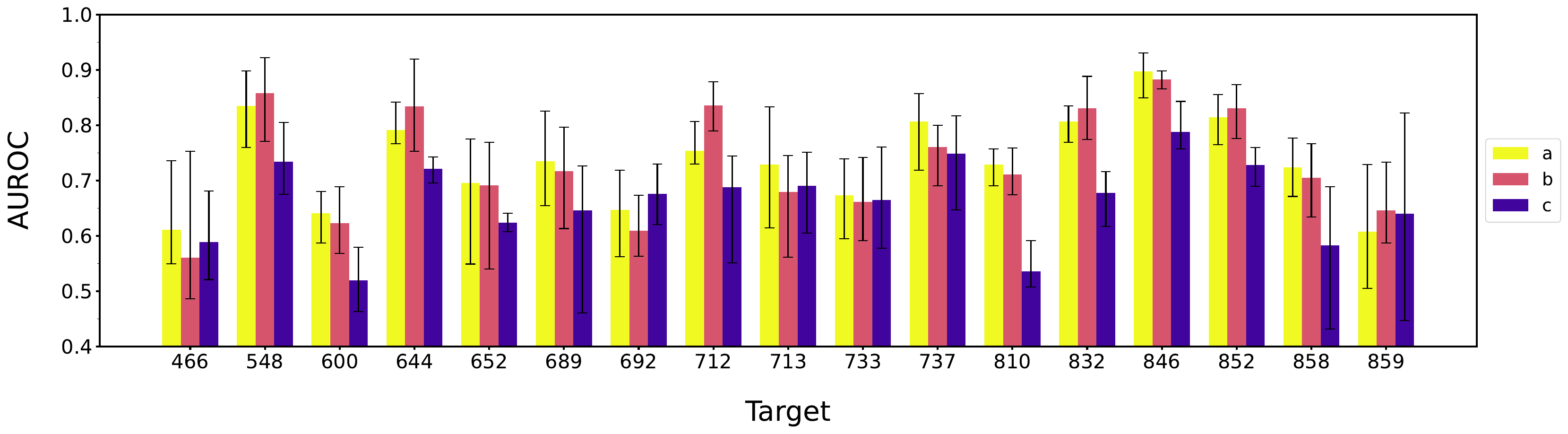}
\caption{The mean AUROC score per MUV target of our classifier trained on Morse features at depth 3 and 32 directions supplemented with the chemical property percentiles (a), chemical property percentiles alone (b), and Morse features at depth 20 and 32 directions (c). The error bars correspond to $95\,\%$ confidence intervals. 
}
\label{fig:muv_auroc_bar}
\end{figure}

\clearpage
We ablate the components of the Morse feature vector, then re-train and test our classifier on the ablated features to determine the relative effects of the components on the generalisation error. 
In Fig.~\ref{fig:dude_diverse_auroc_bar} and \ref{fig:muv_auroc_bar} we compare the per target performance of the chemistry-supplemented Morse features, plain Morse features and chemical percentile features. 
For most targets, chemical percentile features have stronger predictive power than plain Morse features.
It is important to note that the chemical components do not purely contain chemical information since their computation requires knowledge of the 2D molecular graph connectivity, which is a form of shape information.


\begin{figure}[!ht]
\centering
\includegraphics[width=0.45\linewidth]{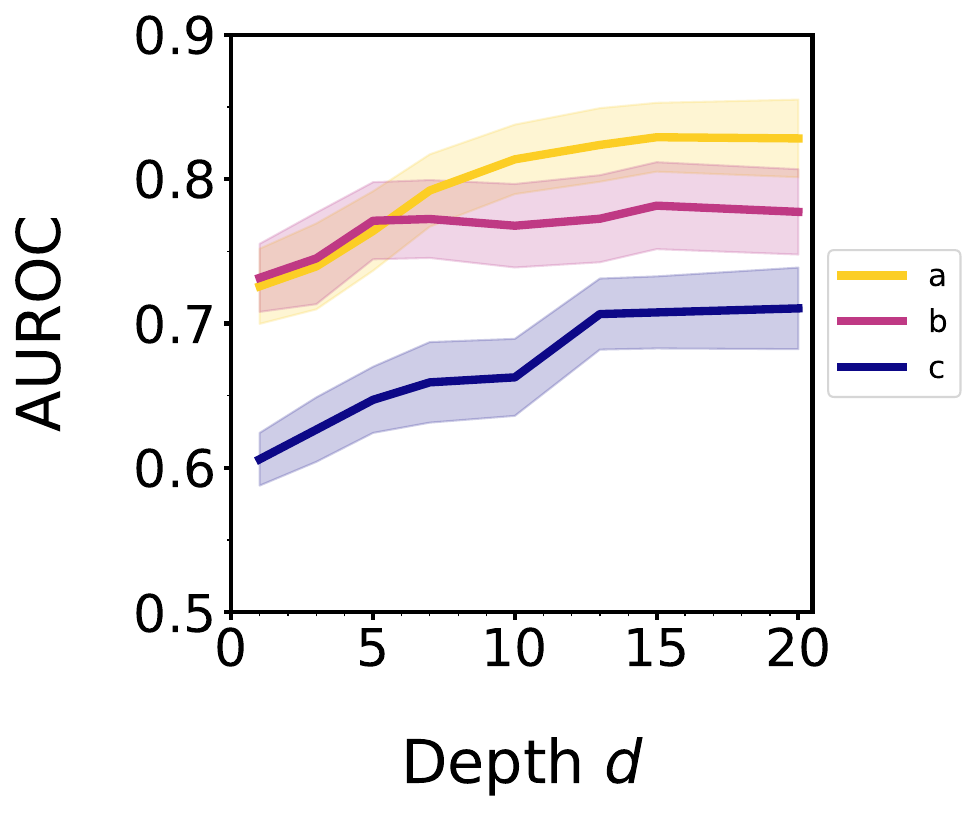}
\caption{The mean AUROC score against depth of variants of the Morse feature vector (32 directions) for the DUD-E Diverse dataset. Here depth refers to the top number (by descending height) of critical vertices in the standard Morse transform (a), the top number (by descending height) of critical \textit{and} regular vertices retained in a variant of the Morse transform (b), the number of randomly selected (without replacement) vertices in another variant of the Morse transform (c).  The error bars correspond to $95\,\%$ confidence intervals. 
}
\label{fig:dude_diverse_auroc_alternate_line}
\end{figure}

\begin{figure}[!ht]
\centering
\includegraphics[width=0.45\linewidth]{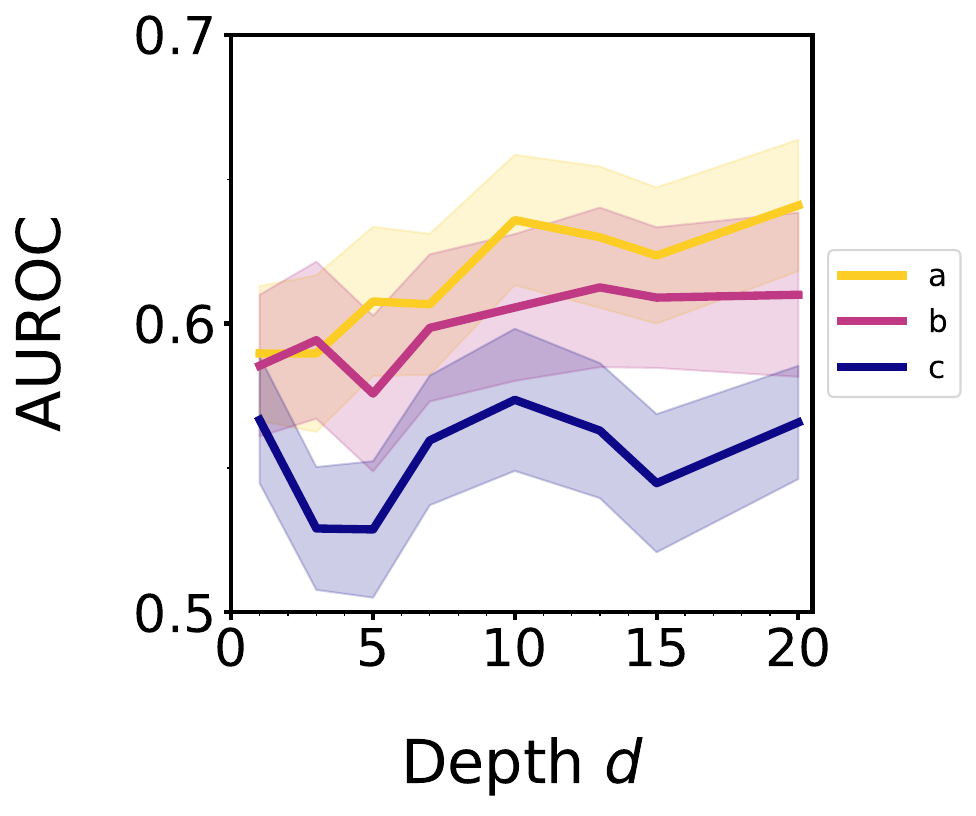}
\caption{The mean AUROC score against depth of variants of the Morse feature vector (32 directions) for the MUV dataset. Here depth refers to the top number (by descending height) of critical vertices in the standard Morse transform (a), the top number (by descending height) of critical \textit{and} regular vertices retained in a variant of the Morse transform (b), the number of randomly selected (without replacement) vertices in another variant of the Morse transform (c).  The error bars correspond to $95\,\%$ confidence intervals. 
}
\label{fig:muv_auroc_alternate_line}
\end{figure}

\clearpage
In Fig.~\ref{fig:dude_diverse_auroc_alternate_line} and \ref{fig:muv_auroc_alternate_line} we see that for both D8 and MUV, at high depth the Morse-theoretic approach of focusing on critical vertices (a) yields better performance than considering both critical and regular vertices together (b), or randomly selecting vertices without replacement (c). At low depth, the performance of the Morse-theoretic approach and approach (b) are similar. Indeed, if depth equals one, then the Morse-theoretic approach and approach (b) are guaranteed to produce the same feature vector as the highest vertex is always critical (it has an empty upper link). As the depth increases and more vertices are considered, then there is a larger chance for the set of highest critical vertices to differ from the set of highest vertices.


\clearpage
\section{Direction sampling robustness}
\label{appendix:direction_robustness}

\begin{figure*}[!ht]
    \centering
    \begin{subfigure}{0.45\linewidth}
    \centering
    \includegraphics[width=\linewidth]{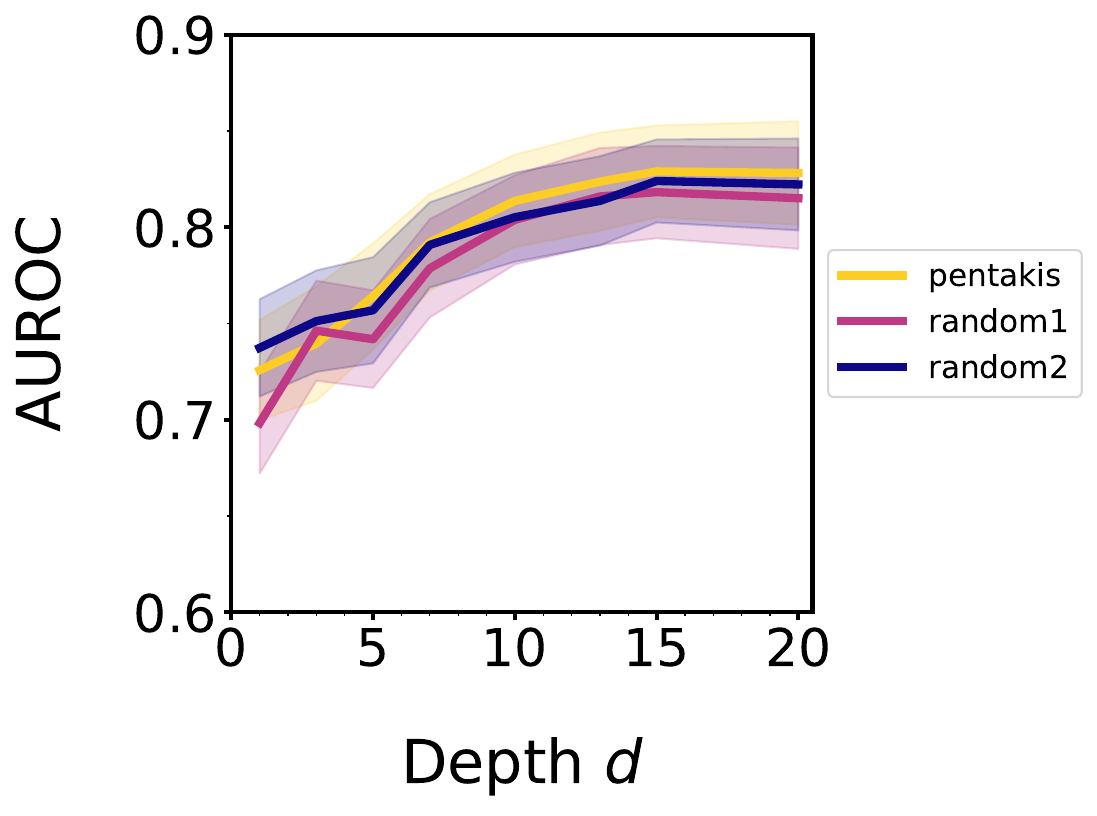}
    \caption{DUD-E}
    \label{subfig:dude_directions}
    \end{subfigure}
    \hspace{5ex}
    \begin{subfigure}{0.45\linewidth}
    \centering
    \includegraphics[width=\linewidth]{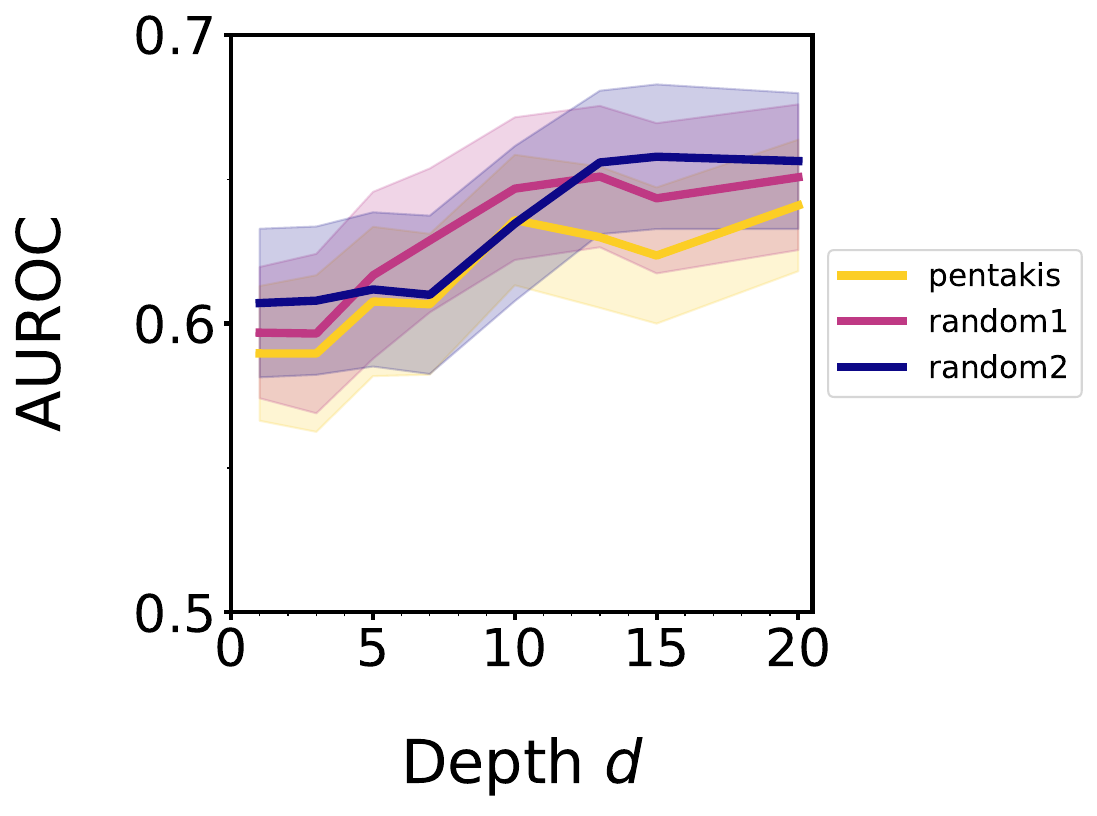}
    \caption{MUV}
    \label{subfig:muv_directions}
    \end{subfigure}\\  
    \caption{Plots of the AUROC of the classifier as a function of the depth of the Morse feature vector for various sets of $32$ directions: one set of directions corresponding to the vertices of the pentakis dodecahedron and two sets corresponding to uniform random samples of the $2$-sphere. The error bars are $95\,\%$ confidence intervals. 
    }
  \label{fig:direction_robustness}
\end{figure*}

In Fig.~\ref{fig:direction_robustness} we plot the AUROC of the classifier trained on Morse features with different sets of directions
with the same cardinality. 
We choose three sets of directions: $32$ pentakis dodecahedron directions and two sets of $32$ directions uniformly randomly sampled  from the $2$-sphere.
For DUD-E, there is a small difference in the performance of the different sets of directions.
For MUV, the difference is greater (most notably at depth $d=15$) but across most depths there is not a large difference. 
Therefore, at least for $32$ directions, the efficacy of Morse features is relatively robust against uniform resampling of the directions, though more samples would be required for statistical significance.

We anticipate that the robustness to uniform resampling of the directions would increase as the number of directions increases but we have not tested this and it would be useful to investigate this in future work.
In general, we would not expect Morse features to be robust to \textit{non-uniform} resampling, for instance if all $32$ directions were resampled from a small solid angle fraction of the $2$-sphere as then the resampled directions would be likely to miss geometrically pertinent views of the shape.

\clearpage
\section{Noise robustness}
\label{appendix:noise}

Morse features were regenerated for modified point clouds from DUD-E and MUV with various amounts of added noise. 
Specifically, zero-mean Gaussian noise at various standard deviations $\sigma \in \{0.0, 0.05, 0.10, 0.20, \allowbreak 0.50 \}\,\text{\AA}$ was added to the co\"{o}rdinates of each atom after the pruned Delaunay triangulation was \linebreak generated.
Molecular conformers are considered to be identical when the root mean square deviation between them is less than $0.10 \,\text{\AA}$ \citep{geom_conformers}; therefore, noise with standard deviation above this threshold, $\sigma > 0.10\,\text{\AA}$, is likely to lead to chemically significant changes to the shape of the molecule. 
Consequently, a shape-based method should be robust to noise up to that threshold (when the molecule can still be considered the same shape) and sensitive to noise after that threshold (when the molecule is likely to have a chemically different shape). It is expected that the Morse feature vector is affected by noise to some extent because it examines critical vertices and the criticality of each vertex depends on its upper link, which is sensitive to changes in the height of each member vertex. 

\begin{figure}[!ht]
    \centering
    \begin{subfigure}{0.45\linewidth}
    \centering
    \includegraphics[width=\linewidth]{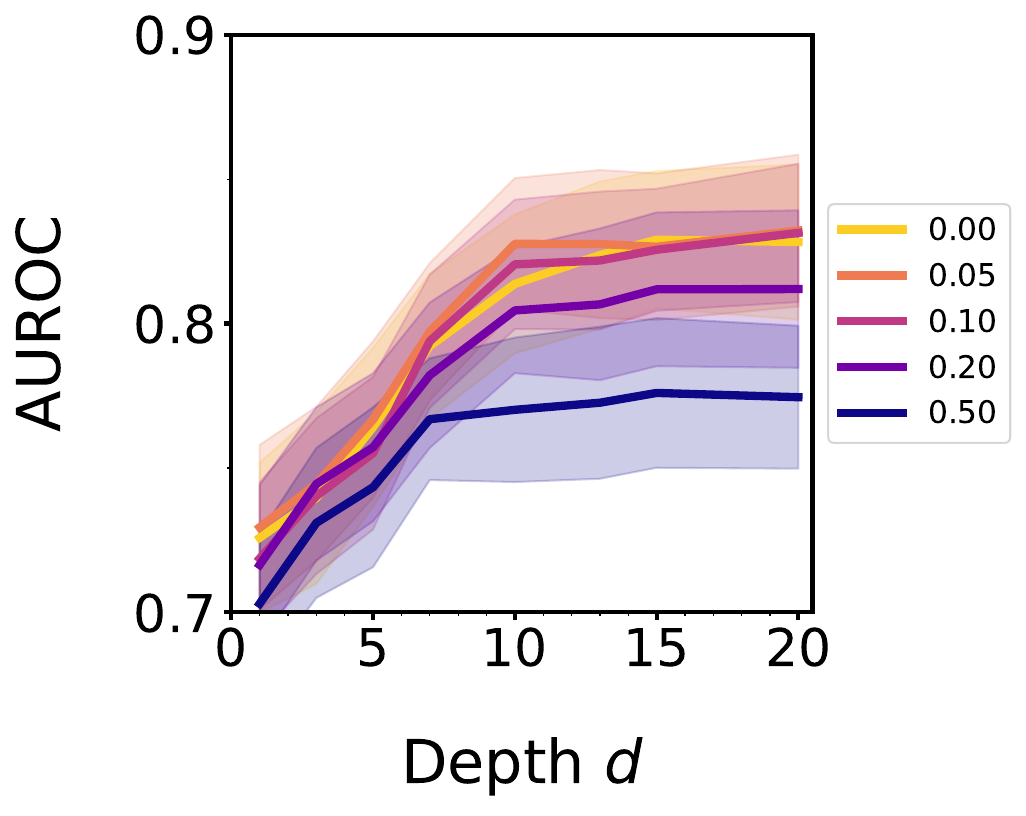}
    \caption{DUD-E}
    \label{subfig:dude_noise}
    \end{subfigure}
    \hspace{5ex}
    \begin{subfigure}{0.45\linewidth}
    \centering
    \includegraphics[width=\linewidth]{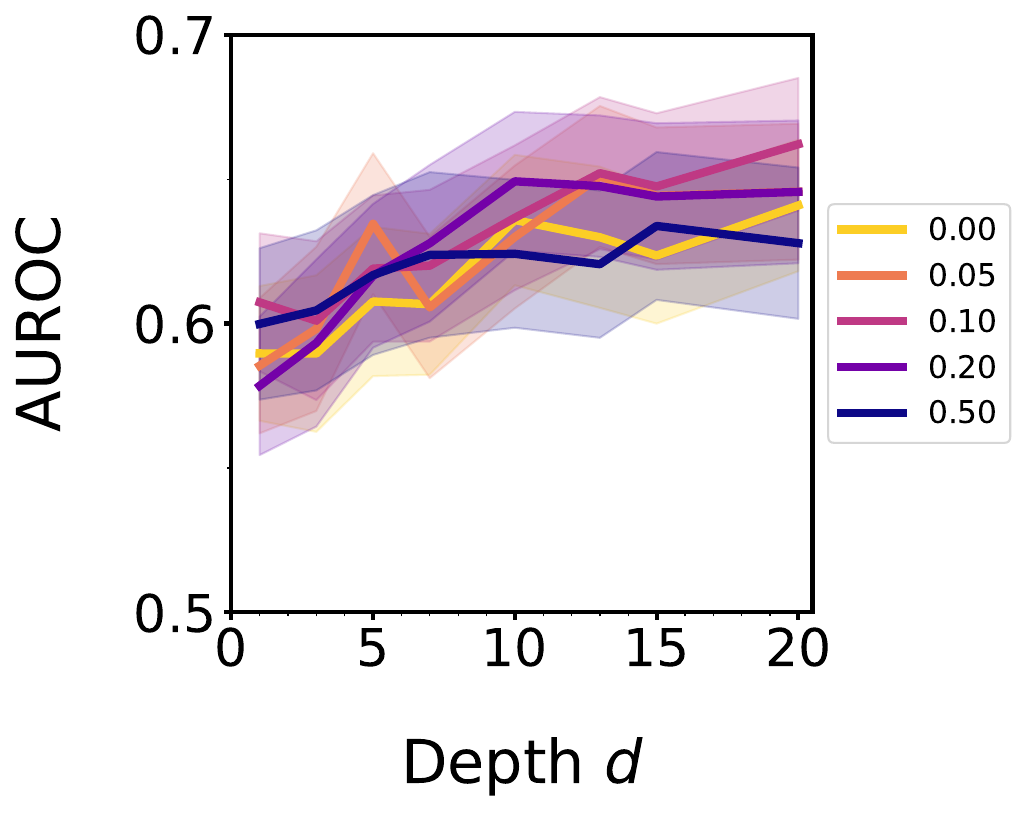}
    \caption{MUV}
    \label{subfig:muv_noise}
    \end{subfigure}\\  
    \caption{Plots of the AUROC of the classifier as a function of the depth of the Morse feature vector for various RMSD of positional Gaussian noise in units of \AA. The error bars correspond to $95\,\%$ confidence intervals. 
    }
  \label{fig:noise_robustness}
\end{figure}

For DUD-E, the performance of the classifier trained on Morse features is not substantially affected until the standard deviation of the noise reaches $\sigma = 0.20\,\text{\AA}$, which demonstrates the desired robustness to positional noise (Fig.~\ref{subfig:dude_noise}). For MUV, the performance of the classifier is relatively unaffected by all levels of noise. It is not clear why this occurs, perhaps the noise has a regularising effect on the classifier 
or the feature is too robust and not sensitive enough or finer shape details are less important in MUV (Fig.~\ref{subfig:muv_noise}). For both datasets, but especially DUD-E, positional noise has a greater effect on the classification of Morse features with higher depth than lower depth. 

\clearpage
\section{Results for additional metrics} 
\label{appendix:other_results}

In Table~\ref{tab:other_metrics} we record the mean scores of various metrics and their standard deviations of our tuned LGBM classifier trained on Morse features at depth 20 and 100 directions (M); chemistry-supplemented Morse features at depth 20 and 100 directions (M+C); 3DP; USR; USRCAT; WHIM; and WHIMu descriptors. Morse features outperform these other shape-based descriptors across all four metrics for the DUD-E Diverse dataset and are either best or joint-best across all metrics for the MUV dataset. Morse features supplemented  with chemical data outperform all these shape- and chemistry-based descriptors across all metrics for both datasets.

\begin{table}[H]
\centering
\caption{The means and standard deviations of performance metric scores of various ligand-based virtual screening methods for the D8 and MUV datasets. For each target the method with the best metric score is displayed in bold and the best shape-based method is underlined. 
}
\label{tab:other_metrics}
\footnotesize
\begin{tabular}{llccccccc}
\toprule 
 & & \multicolumn{4}{c}{Shape} & \multicolumn{3}{c}{Shape \& Chemistry}\\
\cmidrule(lr){3-6} \cmidrule(lr){7-9}
Dataset & Metric & M & 3DP & USR & WHIMu & M+C & USRCAT & WHIM\\
\midrule
\multirow{8.75}{*}{D8} & AUROC & \makecell{$\underline{0.84}$ \\$(0.09)$} & \makecell{$0.62$\\$(0.08)$} & \makecell{$0.76$\\$(0.10)$} & \makecell{$0.81$\\$(0.11)$} & \makecell{${\bf 0.97}$\\$(0.03)$} & \makecell{$0.92$\\$(0.07)$} & \makecell{$0.88$\\$(0.08)$}\\
\addlinespace

& BEDROC\textsubscript{20} & \makecell{$\underline{0.44}$\\$(0.15)$} & \makecell{$0.12$\\$(0.05)$} & \makecell{$0.30$\\$(0.14)$} & \makecell{$0.38$\\$(0.15)$} & \makecell{${\bf 0.85}$\\$(0.13)$} & \makecell{$0.70$\\$(0.18)$} & \makecell{$0.57$\\$(0.17)$}\\
\addlinespace

& EF\textsubscript{1\%} & \makecell{$\underline{19}$\\$(10)$} & \makecell{$3$\\$(3)$} & \makecell{$11$\\$(8)$} & \makecell{$16$\\$(9)$} & \makecell{${\bf 57}$\\$(14)$} & \makecell{$44$\\$(16)$} & \makecell{$30$\\$(12)$}\\
\addlinespace

& REF\textsubscript{1\%} & \makecell{$\underline{0.31}$\\$(0.17)$} & \makecell{$0.04$\\$(0.05)$} & \makecell{$0.18$\\$(0.13)$} & \makecell{$0.26$\\$(0.15)$} & \makecell{${\bf 0.88}$\\$(0.16)$} & \makecell{$0.68$\\$(0.22)$} & \makecell{$0.48$\\$(0.20)$}\\
\midrule
\multirow{8.75}{*}{MUV} & AUROC & \makecell{$\underline{0.66}$\\$(0.12)$} & \makecell{$0.53$\\$(0.10)$} & \makecell{$0.56$\\$(0.12)$} & \makecell{$0.61$\\$(0.13)$} & \makecell{${\bf 0.74}$\\$(0.12)$} & \makecell{$0.63$\\$(0.15)$} & \makecell{$0.64$\\$(0.14)$}\\
\addlinespace

& BEDROC\textsubscript{20} & \makecell{$\underline{0.12}$\\$(0.09)$} & \makecell{$0.06$\\$(0.07)$} & \makecell{$0.07$\\$(0.08)$} & \makecell{$0.10$\\$(0.09)$} & \makecell{${\bf 0.23}$\\$(0.16)$} & \makecell{$0.13$\\$(0.13)$} & \makecell{$0.12$\\$(0.11)$}\\
\addlinespace

& EF\textsubscript{1\%} & \makecell{$\underline{3}$\\$(7)$} & \makecell{$1$\\$(4)$} & \makecell{$2$\\$(6)$} & \makecell{$\underline{3}$\\$(7)$} & \makecell{${\bf 11}$\\$(14)$} & \makecell{$4$\\$(8)$} & \makecell{$3$\\$(7)$}\\
\addlinespace

& REF\textsubscript{1\%} & \makecell{$\underline{0.006}$\\$(0.015)$} & \makecell{$0.002$\\$(0.008)$} & \makecell{$0.004$\\$(0.012)$} & \makecell{$0.005$\\$(0.013)$} & \makecell{${\bf 0.021}$\\$(0.028)$} & \makecell{$0.007$\\$(0.015)$} & \makecell{$0.006$\\$(0.015)$}\\
\bottomrule
\bottomrule
\end{tabular}
\end{table}

\section{Implementation}
\label{appendix:implementation}

Our pipeline is coded in \textsc{python 3}. We compute the WDTs and Betti numbers using the \textsc{gudhi} library \citep{gudhi}; chemical properties with \textsc{rdk}{\small it} \citep{rdkit}; machine learning with \textsc{scikit-learn} \citep{scikit-learn} and \textsc{l}{\small ight}\textsc{gbm} \citep{lightgbm}; hyperparameter tuning with \textsc{ray tune} \citep{raytune}; and throughout we use \textsc{matplotlib} \citep{matplotlib}, \textsc{n}{\small um}\textsc{p}{\small y} \citep{numpy}, \textsc{pandas} \citep{pandas} and \textsc{plotly} \citep{plotly}.

\section{Computational resources}
\label{appendix:computational}

All features were computed on a cluster node consisting of two Intel Platinum 8628 CPUs (a 24 core 2.90\,GHz Cascade Lake CPU) and 384\,GB of memory.

\end{document}